\documentclass[aps,amssymb,amsmath,
twocolumn,showpacs,superscriptaddress,
groupedaddress,ifmbe]{revtex4-1}

\usepackage{graphicx}
\usepackage{dcolumn}
\usepackage{bm}
\usepackage[table,xcdraw]{xcolor}
\usepackage{nth}
\usepackage[ruled,vlined]{algorithm2e}
\usepackage{framed}
\usepackage{listings}
\usepackage[titletoc,title]{appendix}
\usepackage{tikz}
\usepackage{lipsum}

\begin{document}
\title{I. Complete and orthonormal sets of exponential-type orbitals\\ 
with noninteger principal quantum numbers.}
\author{A. Ba{\u g}c{\i}}
\email{abagci@pau.edu.tr}
\affiliation{
Instituto de Modelado e Innovaci{\'o}n Tecnol{\'o}gica (IMIT), Facultad de Ciencias Exactas,
Naturales y Agrimensura, Universidad Nacional del Nordeste, Avda. Libertad 5460, W3404AAS, Corrientes, Argentina
\\
Computational Physics Laboratory, Department of Physics, Faculty of Sciences, Pamukkale University, {\c C}amlaraltı, K{\i}n{\i}kl{\i} Campus, Denizli, Turkey}

\author{P. E. Hoggan}
\affiliation{Institute Pascal, UMR 6602 CNRS, Clermont-Auvergne University, 24 avenue des Landais BP 80026, 63178 Aubiere Cedex, France}
\begin{abstract}
The definition for the Slater$-$type orbitals is generalized. Transformation between an orthonormal basis function and the Slater$-$type orbital with non$-$integer principal quantum numbers is investigated. Analytical expressions for the linear combination coefficients are derived. In order to test the accuracy of the formulas, the numerical Gram$-$Schmidt procedure is performed for the non$-$integer Slater$-$type orbitals. A closed form expression for the orthogonalized Slater$-$type orbitals is achieved. It is used to generalize complete orthonormal sets of exponential$-$type orbitals obtained by Guseinov in [Int. J. Quant. Chem. \textbf{90}, 114 (2002)] to non$-$integer values of principal quantum numbers. Riemann$-$Liouville type fractional calculus operators are considered to be use in atomic and molecular physics. It is shown that the relativistic molecular auxiliary functions and their analytical solutions for positive real values of parameters on arbitrary range are the natural Riemann$-$Liouville type fractional operators.
\begin{description}
\item[Keywords]
Gram-Schmidt method, exponential$-$type orbitals, non$-$integer principal quantum numbers, Riemann$-$Liouville integral
\item[PACS numbers]
... .
\end{description}
\end{abstract}
\maketitle
\section{Introduction} \label{introduction}
Basis functions to be used in the algebraic solution of the Dirac equation for point$-$like nuclei are constructed from the Slater$-$type orbitals ($n^*-$STOs) with non$-$integer principal quantum numbers $\left(n^*, n^* \in \mathbb{R}/\mathbb{N}_{0} \right)$ \cite{1_Slater_1930,2_Infeld_1951, 3_Parr_1957},
\begin{multline}\label{eq:1}
\eta_{n^{*}lm}\left(\zeta, \vec{r}\right)=R_{n^{*}}\left(\zeta, r\right)S_{lm}\left(\theta, \varphi\right)\\
=N_{n^{*}}\left(\zeta\right)r^{n^*-1}e^{-\zeta r}S_{lm}\left(\theta, \varphi\right)
\end{multline}
The functions $S_{lm}$ are normalized complex $\left(S_{lm} \equiv Y_{lm}, Y^{*}_{lm}=Y_{l-m} \right)$ \cite{4_Condon_1935} or real spherical harmonics. $\zeta$ are orbital parameters. The symbol $\mathbb{R}$, $\mathbb{N}_{0}$ are used to denote the sets real and natural numbers. The $n^*-$STOs basis are also promise better results in calculations for atoms \cite{5_Koga_1997, 6_Koga_1997, 7_Koga_1997, 8_Koga_1998, 9_Koga_2000} using the Hartree$-$Fock method because they are simple but more flexible than the $n-$STOs with integer values of $n$, $ \left( n=n^* \Leftrightarrow n^* \in \mathbb{N}_{0} \right)$. Examining advantages of using these orbitals in calculations for molecules on the other hand, has not been possible so far. This is because, the molecular integral evaluation problem for $n^*-$STO had been thought to be nearly insurmountable. The authors published two series of papers that provide both numerical and analytical solutions to this problem. In the first series \cite{10_Bagci_2014, 11_Bagci_2015, 12_Bagci_2015}, numerical approximations were used to produce benchmark values for the two$-$ \cite{10_Bagci_2014, 11_Bagci_2015}, three$-$center \cite{12_Bagci_2015} one$-$ and two$-$electron integrals. The numerical global$-$adaptive scheme with the Gauss$-$Kronrod numerical integration extension was used to obtain the required precision. In the second series, a new fully analytical method of calculation was developed \cite{13_Bagci_2018,14_Bagci_2018, 15_Bagci_2020}. The molecular integrals were expressed through relativistic molecular auxiliary functions defined in prolate spheroidal coordinates. The methods so far presented by the authors are useful in calculation up to three$-$center integrals.\\
A solution for the four$-$center problem was suggested by Allouche in \cite{16_Allouche_1976} using the numerically-adaptive transformation for radial part of $n^*-$STO. The translation from one$-$center to another given as,
\begin{multline}\label{eq:2}
r_{B}^{n^{*}}e^{-\zeta r_{b}}
\\
=\sqrt{2\pi}\sum_{l=0}^{\infty}\dfrac{1}{r_{A}R_{AB}}V_{n^{*}+2l}\left(r_{A},R_{AB},\zeta \right)Y_{l0}\left(\theta_{A}, \varphi_{A}\right)
\end{multline}
\begin{multline}\label{eq:3}
V_{n^{*}l}\left(r_{A},R_{AB},\zeta\right) \\
=\int_{\vert R_{AB}-r_{A} \vert}^{\vert R_{AB}+r_{A} \vert}
r^{n^{*}}_{B}e^{-\zeta r}\mathcal{P}_{l0}\left(Cos \theta_{A} \right)dr_{B}
\end{multline}
$\mathcal{P}_{lm}$ are the normalized associated Legendre polynomials.\\
Four$-$center integrals delineate an analytical problem that is the last bottleneck impeding us from effectively using the $n^*-$STO in both relativistic and non$-$relativistic molecular calculations. It is considered rightly as extremely arduous since the $n^*-$STO do not have infinite series representation formulas i.e., power series for a function such as $z^\rho, z\in \mathbb{C}$, and $\rho=\mathbb{R}/\mathbb{N}_{0}$ are not analytic at the origin (see \cite{15_Bagci_2020} and references therein). The symbol $\mathbb{C}$, here is used to denotes the sets of complex numbers. If, on the other hand, a complete orthonormal set of exponential functions with non$-$integer principal quantum numbers were introduced, then we may overcome such difficulty. 
\begin{table}
\caption{Possible variants for an electron configuration depending to the values quantum numbers for $n>0$.}
\label{table:V1n1}
\begin{ruledtabular}
\begin{tabular}{cccc}
Variants & $l^{*}$ & $n^{*}$ & $m^{*}$\\
\hline
$n\geq 1$& &  &\\
$\nth{1}$ & $l^{*} \in \mathbb{N}_{0}$ & $n^{*} \geq l^{*}+1$ & $-l^{*} \leq m^{*} \leq l^{*}$ \\
$\nth{2} \left(a\right)$ & $l^{*} \in \mathbb{R}^{+}, l^{*} \geq l+\epsilon_{1}$ & $n^{*} \geq l^{*}+1$ & $-l \leq m^{*} \leq l$ \\
$\nth{2} \left(b\right)$ & $l^{*} \in \mathbb{R}^{+}, l^{*}\geq l+\epsilon_{1}$ & $n^{*}\geq l^{*}+1$ & $-l^{*} \leq m^{*} \leq l^{*}$\\
$\nth{3} \left(a\right)$ & $l^{*} \in \mathbb{R}^{+}, l^{*}\geq l\epsilon_{1}\pm \epsilon_{2}$ & $n^{*}\geq l^{*}+1$ & $-l \leq m^{*} \leq l$\\
$\nth{3} \left(b\right)$ & $l^{*} \in \mathbb{R}^{+}, l^{*}\geq l\epsilon_{1}\pm \epsilon_{2}$ & $n^{*}\geq l^{*}+1$ & $-l^{*} \leq m^{*} \leq l^{*}$\\
$n>0$& &  &\\
$\nth{4} \left(a\right)$ & $l^{*} \in \mathbb{R}^{+}, l^{*}\geq l\epsilon_{1}\pm \epsilon_{2}$ & $n^{*}\geq l^{*}+\epsilon_{3}$ & $-l \leq m^{*} \leq l$\\
$\nth{4} \left(b\right)$ & $l^{*} \in \mathbb{R}^{+}, l^{*}\geq l\epsilon_{1} \pm \epsilon_{2}$ & $n^{*}\geq l^{*}+\epsilon_{3}$ & $-l^{*} \leq m^{*} \leq l^{*}$
\end{tabular}
\footnotetext[1]{$l=\lfloor l^{*} \rfloor=0,1,2,3...,$, $0 < \epsilon_{1} \leq 1$, $\epsilon_{2}$ is shifting parameter $0 \leq \epsilon_{2}\leq \epsilon_{1}$. The steps $\epsilon_{1}, \epsilon_{3}$ are used to determine elements of a sequence.}
\end{ruledtabular}
\end{table}
In this work, possible variants of electron configurations that may represent the set of complete orthonormal functions for non$-$integer order is investigated. The definition given in the Eq. (\ref{eq:1}) for the Slater$-$type orbitals are generalized, accordingly. Such generalization exhibits the form of Laguerre functions that the $n^*-$STOs are obtained by their simplifications. A complete set of orthonormal functions of non$-$integer order is derived in following. Closed form expressions are obtained for the expansion and translation coefficients from one point to another between $\eta_{n^{*}lm}$ and orthonormalized $n^*-$STO $\left(\chi_{n^{*}lm} \right)$ or visa versa. The resulting expressions are compaired with ones obtained from the numerical Gram$-$Schmidt ortho$-$gonalization procedure \cite{17_Werneth_2010}. These relationships are then used to generalize the $\chi_{n^{*}lm}$ to Guseinov's complete orthonormal sets of exponential orbitals ($\Psi_{n^{*}l^{*}m^{*}}^{\alpha \epsilon}-$ETOs) \cite{18_Guseinov_2002} but this time with non$-$integer values of principal and angular momentum quantum numbers. Note that, Guseinov in \cite{18_Guseinov_2002} followed a similar route to generalize his early work on the orthonormalized $n-$STO $\left(\chi_{nlm} \right)$ \cite{19_Guseinov_1980}. The present work deals with a harder problem since the quantum numbers represent the electron configuration are not anymore fixed, i.e., they are not in set of positive integer numbers.

\textit{Dynamic quantum numbers} in an electron configuration brings about new discussions. The $\Psi_{n^{*}l^{*}m^{*}}^{\alpha \epsilon}-$ETOs now, involve generalized Laguerre polynomials \cite{20_Polya_1976} and Legendre polynomials \cite{21_Hobson_1931} of non$-$integer order. They are calculated through the Riemann$-$Liouville integrals \cite{22_Oldham_1974} referred to as fractional calculus operators \cite{22_Oldham_1974,23_Kilbas_2006},
\begin{align}\label{eq:4}
I^{\mu}f\left(x\right)
=\frac{1}{\Gamma\left(\mu\right)}
\int_{0}^{x}f\left(t\right)\left(x-t\right)^{\mu-1}dt,
\end{align}
\begin{align}\label{eq:5}
D^{\mu}f\left(x\right)
=\frac{1}{\Gamma\left(1-\mu\right)}
\frac{d}{dx}\int_{0}^{x}f\left(t\right)\left(x-t\right)^{-\mu}dt,
\end{align}
for smallest integer that $r$ exceeds $\mu$,
\begin{align}\label{eq:6}
D^{\mu}f\left(x\right)
=\frac{1}{\Gamma\left(r-\mu\right)}
\frac{d^{r}}{dx^{r}}\int_{0}^{x}f\left(t\right)\left(x-t\right)^{r-\mu-1}dt.
\end{align}
Note that, Riemann$-$Liouville type fractional calculus operators involve special functions such as beta and hypergemetric functions are also derived \cite{24_Ozarslan_2010,25_Luo_2013,26_Luo_2014,27_Nisar_2020,28_Jain_2022}.\\
The problem in the factional operators is that they are too much complicated to be applied in the quantum mechanics. This is because solving a differential equation require either use of the Laplace transform or fractional power series technique \cite{23_Kilbas_2006}. Once again, non$-$analyticity of power series remains as a serious hindrance in solving an integral represented via the fractional calculus operators. The studied \cite{24_Ozarslan_2010,25_Luo_2013,26_Luo_2014,27_Nisar_2020,28_Jain_2022} so far have carried out within the limit of the parameters which ensure the convergence for power series. Let us consider to derive an analytical relationship for an operator involve the power functions with non$-$integer exponent. The binomial expansion for such a power function is,
\begin{align}\label{eq:7}
\left(x+y\right)^{p}
=\sum_{k=0}^{\infty}F_{k}\left(p\right)x^{k}y^{p-k},
\end{align}
where $F_{q}\left(p\right)$ represents the binomial coefficients, $F_{q}\left(p\right)=\binom{p}{q}$. This series converges for $p \geq 0$ an integer, or $\vert x/y \vert<1$. The convergence of an analytical relationship to be derived for a fractional calculus operator involve a special function such as gamma, beta or hypergeometric functions, is much more restricted in terms of the range for values of parameters. The convergence for any representation of a special function should be investigated in the well$-$defined domains. In general the special functions naturally involve large number of parameters. Each representation (recurrence relationship, continued fraction, series expansion formula or asymptotic method) for the special functions depending on the values of parameters construct a different domain of convergence.\\
Four domains with different representations were established following a compromise between efficiency and accuracy for instance in \cite{29_Gill_2012} for the incomplete gamma functions:
\begin{align}\label{eq:8}
\gamma\left(a,x\right)=\int_{0}^{x}t^{a-1}e^{-t}dt,
\end{align}
\begin{align}\label{eq:9}
\Gamma\left(a,x\right)=\int_{x}^{\infty}t^{a-1}e^{-t}dt,
\end{align}
and,
\begin{align}\label{eq:10}
P\left[a,x\right]=\frac{1}{\Gamma\left(a\right)}\gamma\left(a,x\right),
\end{align}
\begin{align}\label{eq:11}
Q\left[a,x\right]=\frac{1}{\Gamma\left(a\right)}\Gamma\left(a,x\right),
\end{align}
normalized incomplete gamma functions.  An approach for computing the special functions aforementioned above without error is still being studied in the literature \cite{29_Gill_2012,30_Bujanda_2017,31_Ansari_2019,32_Reynolds_2021,33_Fejzullahu_2021,34_Pearson_2017}. The hypergeometric functions suffer with the same problem as well. Enormous number of papers have been published for efficient computation of the hypergeometric functions \cite{33_Fejzullahu_2021,34_Pearson_2017} (see also references therein). The well$-$known definition for instance, for the Gauss hypergeometric functions within unit disk $\vert x \vert <1$ is given by,
\begin{align}\label{eq:12}
_{2}F_{1}\left[a,b;c;x\right]
=\sum_{k=0}^{\infty}\frac{\left(a\right)_{k}\left(b\right)_{k}}{\left(c\right)_{k}}
\frac{z^{k}}{k!}.
\end{align}
Some values (of interest) for parameters in a special function arising in the applications of $n^{*}-$STOs for molecules are generally out of the range that ensures the convergence in an expansion representation via the power series \cite{11_Bagci_2015,13_Bagci_2018,14_Bagci_2018,15_Bagci_2020}. The authors have also overcome this serious bottleneck in \cite{11_Bagci_2015,13_Bagci_2018,14_Bagci_2018,15_Bagci_2020,35_Bagci_2020,36_Bagci_2022} using the standard calculus and its operators while evaluating relativistic molecular auxiliary functions,
\begin{multline} \label{eq:13}
\left\lbrace \begin{array}{cc}
\mathcal{P}^{n_1,q}_{n_{2}n_{3}n_{4}}\left(p_{123} \right)
\\
\mathcal{Q}^{n_1,q}_{n_{2}n_{3}n_{4}}\left(p_{123} \right)
\end{array} \right\rbrace\\
=\frac{p_{1}^{\sl n_{1}}}{\left({\sl n_{4}}-{\sl n_{1}} \right)_{\sl n_{1}}}
 \int_{1}^{\infty}\int_{-1}^{1}{\left(\xi\nu \right)^{q}\left(\xi+\nu \right)^{\sl n_{2}}\left(\xi-\nu \right)^{\sl n_{3}}}\\ \times
\left\lbrace \begin{array}{cc}
P\left[{\sl n_{4}-n_{1}},p_{1}\mathfrak{f}_{ij}^{k}\left(\xi, \nu\right) \right]
\\
Q\left[{\sl n_{4}-n_{1}},p_{1}\mathfrak{f}_{ij}^{k}\left(\xi, \nu\right) \right]
\end{array} \right\rbrace
e^{p_{2}\xi-p_{3}\nu}d\xi d\nu,
\end{multline}
where,
\begin{align}\label{eq:14}
\mathfrak{f}_{ij}^{k}\left(\xi, \mu \right)
=\left(\xi \mu\right)^{k}\left(\xi+\nu \right)^{i}\left(\xi-\nu\right)^{j},
\end{align}
stands to represent the elements required to generate a potential. For Coulomb potential it has a form that, $i=1$, $k=j=0$; $\mathfrak{f}_{10}^{0}\left(\xi, \nu\right)=\left(\xi+\nu \right)$.
\begin{figure*}[htp!]
    \centering
    \includegraphics[width=1.0\textwidth]{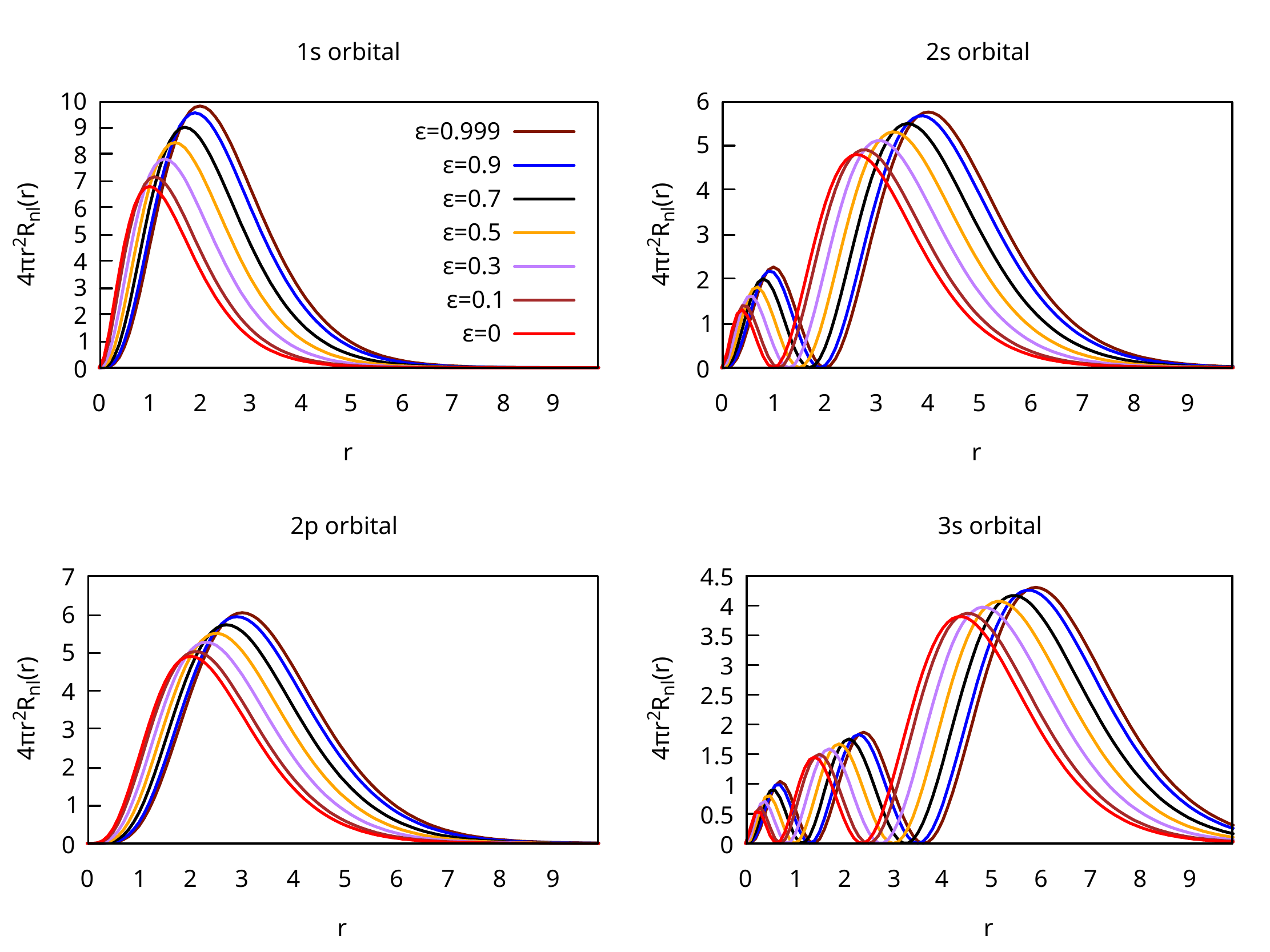}
    \caption{Radial distribution functions $\left\lbrace 4\pi r^{2}\left[R^{l^{*}}_{n^{*}}\left(\zeta, r\right)\right]^{*} R^{l^{*}}_{n^{*}}\left(\zeta, r\right) \right\rbrace$ for different values of $\epsilon$ and in atomic units $\left(a.u.\right)$. They are obtanied through the Eq. (\ref{eq:17}) where $\zeta=1$. $\epsilon=0$ coresponds to Coulomb$-$Sturmian functions.}
    \label{fig:infeld1}
\end{figure*}
The relation between the Riemann$-$Liouville type fractional calculus operators and relativistic molecular auxiliary functions is examined in the present study. It is shown that, analytical formulas derived for solutions of molecular auxiliary functions are useful in fractional calculus. Although, it is specified in the relevant parts of the paper an asterix as a superscript on a parameter or a vriable indicates that this parameter or variable is in sets of real numbers.
\section{On the Schr{\"o}dinger$-$like radial differential equation and its properties}\label{sec:SchRadPro}
The $n-$STOs,
\begin{align}\label{eq:15}
\chi_{nlm}^{n-1}\left(\zeta, \vec{r}\right)=\eta_{nlm}\left(\zeta, \vec{r}\right)
\end{align}
where, $n=1,2,3,... , 0\leq l \leq n-1, -l \leq m \leq l$, are obtained by simplification of Laguerre functions in hydrogen-like orbitals $\left(\chi_{nlm} \right)$ by keeping only the term of the highest power of $r$. They are complete in Sobolev space [$W_{2}^{1}\left(\mathbb{R}^{3} \right)$, which is a proper subspace of the Hilbert space $L^{2}\left(\mathbb{R}^{3} \right)$] \cite{37_Klahn_1977}. The sequence for orbital parameters $\zeta$ is a Cauchy sequence and it has an accumulation point with respect to elements $\left(\eta_{nlm}\right)$ of space. The definition given in the Eq. (\ref{eq:1}) for $n^{*}-$STOs suggested by Parr and Joy \cite{3_Parr_1957} following Infeld's work \cite{2_Infeld_1951} on the other hand, is insufficient. It does not explicitly indicate from which orthonormal function set that satisfies $\langle \chi_{n^{*}} \vert \chi_{{n'}^{*}}\rangle=\delta_{{n^{*}}{{n'}^{*}}}$, it is derived. Infeld in order to derive the $n^{*}-$STOs, suggested a more generic form for radial part of the Schr{\"o}dinger equation:
\begin{multline}\label{eq:16}
\dfrac{d^{2}R\left(r\right)}{dr^{2}}+\dfrac{2}{R\left(r\right)}-\dfrac{(l^{*})(l^{*}+1)}{r^2}R\left(r\right)
\\
-\dfrac{1}{(n^{*})^2}R\left(r\right)=0.
\end{multline}
He replaced the principal and angular momentum quantum numbers by $n^{*}=n+\epsilon$, $l^{*}=l+\epsilon$ so that the difference $n^{*}-l^{*}-1 \in \mathbb{N}_{0}$ continues to hold. Here, $\epsilon$, $0 \leq \epsilon < 1$ is a constant. Infeld, instead of presenting an explicit solution he preferred to use his factorization method to produce a solution via recurrence relationships based on the $n^{*}-$STOs. The explicit solution for the Eq. (\ref{eq:16}) which satisfies the boundary condition imposed on $R\left(r\right)$ at $r=0$ is determined as,
\begin{multline}\label{eq:17}
R_{n^{*}}^{l^{*}}\left(\zeta, r\right)\\
=N_{n^{*}l^{*}}\left(\zeta\right)
(2\zeta r)^{l^{*}}e^{-\zeta r}
\mathcal{L}_{n^{*}-l^{*}-1}^{2l^{*}+1}\left(2\zeta r\right),
\end{multline}
\begin{align}\label{eq:18}
N_{n^{*}l}\left(\zeta\right)
=\Bigg[
\dfrac{\left(2\zeta\right)^{3}\Gamma \left( n^{*}-l^{*} \right)}{2n^{*}\Gamma\left( n^{*}+l^{*}+1\right)}
\Bigg]^{1/2},
\end{align}
here, $\zeta=\frac{1}{n^{*}}$. $\mathcal{L}_{q-p}^{p}\left(z\right)$ are associated Laguerre polynomials. $p$ not necessarily to be integer and in this case $\mathcal{L}_{q-p}^{p}\left(z\right)$ is called to as "generalized" Laguerre polynomials \cite{38_Magnus_1966}. Note that the definition is given here with quotation marks because the classification for Laguerre functions is the subject of a next section. If $\epsilon=0$, the solution reduces to the standard Schr{\"o}dinger equation. For $\zeta$, $\zeta>0$, the functions given in the Eq. (\ref{eq:17}) satisfy the following orthonormality relationship,
\begin{align}\label{eq:19}
\int
R^{l^{*}}_{n^{*}}\left(\zeta, r\right)
\left(\dfrac{{n'}^{*}}{\zeta r}\right)
R^{l^{*}}_{{n'}^{*}}\left(\zeta, r\right)dV
=\delta_{n^{*}{n'}^{*}}.
\end{align}
The Slater$-$type orbitals with noninteger principal quantum numbers were obtained by making a similar simplification for Laguerre polynomials in the Eq. (\ref{eq:17}) but only for $n^{*}$, $n^{*} \geq 1$. They are defined as,
\begin{align}\label{eq:20}
\eta_{n^{*}l^{*}m^{*}}\left(\zeta, \vec{r}\right)
=R_{n^{*}}^{n^{*}-1}\left(\zeta, r\right)S_{l^{*}m^{*}}\left(\theta, \varphi\right),
\end{align}
where,
\begin{align}\label{eq:21}
R_{n^{*}}^{n^{*}-1}\left(\zeta, r\right)
=N_{n^{*}}\left(\zeta\right)r^{n^*-1}e^{-\zeta r}
\end{align}
First difference between Eq. (\ref{eq:1}) and Eq. (\ref{eq:21}) is that Parr and Joy in the Eq. (\ref{eq:1}) advocated to use the $n^{*}-$STOs with variationally determined noninteger values of $n^{*}$, $n^{*} >0$. Therefore, $R_{n^{*}}\left(\zeta, r\right) = R_{n^{*}}^{n^{*}-1}\left(\zeta, r\right)$ only if $n^{*} \geq l^{*}+1$. The second is that in the Eq. (\ref{eq:21}) the angular momentum quantum numbers $l^{*}, l^{*} \in \mathbb{R}^{+}$. The associated Legendre polynomials constitute the spherical harmonics are now of degree $l^{*}$ and orders $-l^{*} \leq m^{*} \leq l^{*}$. They rigorously were analyzed by Lebedev in \cite{39_Lebedev_1965} following Hobson \cite{21_Hobson_1931}. The most recent notation as $R_{n^{*}l^{*}} \equiv R^{l^{*}}_{n^{*}}$ is used in the next sections.

There exist infinite number of linearly independent sequences that can be constructed from the Eq. (\ref{eq:21}). They are chosen depending on the values of $\epsilon$. Infinite number of orthonormal bases in Hilbert space accordingly, can numerically be produced, via the Gram$-$Schmidt method \cite{17_Werneth_2010} (see Appendix \ref{sec:appendicesa} for details). All of those for $n^{*} \geq 1$ are achieved from the Eq. (\ref{eq:17}). A complete and orthonormal set of vector in Hilbert space is obtained from three possible variants of quantum numbers that represent an electron configuration for $n^{*} \geq 1$ (Table \ref{table:V1n1}). These variants are generated according to the values for angular momentum quantum number $l^{*}$ and in all $n^{*}-l^{*}-1 \in \mathbb{N}_{0}$ still holds. Note that, solutions for the Eq. (\ref{eq:16}) such as for $n=0$ which have no counterpart in the ordinary central field problem are not considered throughout this work.

The "generalized" Laguerre polynomials $\mathcal{L}^{p^{*}}_{q^{*}-p^{*}}$ with $q^{*}-p^{*} \in \mathbb{N}_{0}$ and $Re\left(p\right)>-1$ are available to be used in a complete orthonormal set for all variants given in the Table \ref{table:V1n1}. They are the solution for the following differential equation,
\begin{multline}\label{eq:22}
x\dfrac{d^{2}y\left(x\right)}{dx^2}+ \left(p^{*}+1-x\right)\dfrac{dy\left(x\right)}{dx}\\
+\left(q^{*}-p^{*}\right)y\left(x\right)=0.
\end{multline}
They possess explicit expression in terms of confluent hypergeometric function of the first kind as, 
\begin{multline}\label{eq:23}
\mathcal{L}^{p^{*}}_{q^{*}-p^{*}}\left(x\right)
\\
=\dfrac{\left(q^{*}+1\right)_{q^{*}-p^{*}}}{\Gamma\left(q^{*}-p^{*}+1\right)}
{_{1}}F_{1}\left[-\left(q^{*}-p^{*}\right);p^{*}+1;x\right].
\end{multline} 
The Rodrigues representation for the "generalized" Laguerre polynomials is,
\begin{align}\label{eq:24}
\mathcal{L}^{p^{*}}_{q^{*}-p^{*}}\left(x\right)
=\dfrac{{e^{x}}x^{-p^{*}}}{\Gamma\left(q^{*}-p^{*}+1\right)}
\dfrac{d^{q^{*}-p^{*}}}{dx^{q^{*}-p^{*}}}\left(e^{-x}x^{q^{*}}\right).
\end{align}
The form of the differential equation given in the Eq. (\ref{eq:22}) remark a significant difference between the associated and "generalized" Laguerre polynomials. The differential equation of the usual Laguerre polynomials $L_{q}\left(x\right)$, $q^{*} \in \mathbb{N}_{0} \Rightarrow q^{*}=q$ is a special case of the Eq. (\ref{eq:22}) only if $p$, $p^{*} \in \mathbb{N}_{0} \Rightarrow p^{*}=p$ is an integer. Only in this case the Rodrigues representation for the usual Laguerre polynomials becomes a special case of the Eq. (\ref{eq:24}). So far it has been known the opposite but, dropping the restriction on $p$, by replacing it with $p^{*}$ removes the coherence between the usual and "generalized" Laguerre polynomials. Spacial case of the Eq. (\ref{eq:24}) for $p^{*}=0$, here is referred to as \textit{transitional Laguerre polynomials},
\begin{align}\label{eq:25}
\mathcal{L}_{q^{*}}\left(x\right)
=\mathcal{L}^{0}_{q^{*}}\left(x\right)
=\dfrac{{e^{x}}}{\Gamma\left(q^{*}+1\right)}
\dfrac{d^{q^{*}}}{dx^{q^{*}}}\left(e^{-x}x^{q^{*}}\right).
\end{align}
They represent a form between standard Laguerre and fractional Laguerre functions. Necessity for existence of the transitional Laguerre polynomials is now obvious due to completeness of their generalized form [Eq. (\ref{eq:24})]. This is true even if $q^{*}-p^{*} \in \mathbb{N}_{0}$. Nevertheless, presence of the Eq. (\ref{eq:25}) form also a reason for $q^{*}-p^{*} \in \mathbb{R}^{+}$ in the Eq. (\ref{eq:24}) and permits finally, for the values of principal quantum numbers $n^{*}$, $n^{*}>0$ in the Eq. (\ref{eq:20}) (Table \ref{table:V1n1}). There must be a Laguerre$-$type differential equation with nonsingular solution for the range of values of quantum numbers given in the Table \ref{table:V1n1}. The form for series solution of such differential equation is accordingly assumed to be as,
\begin{align}\label{eq:26}
y=\sum_{r=0}^{\infty} a_{r}x^{\left(r+s\right)\epsilon+\epsilon'}.
\end{align}
The discussions on the evaluation of the orthogonal polynomials with noninteger orders will be subject of following sections.
\section{On the Infeld's Complete and orthonormal functions with non$-$integer principal quantum numbers}\label{sec:infeld}
\subsection{Generalizing the Infeld's Schr{\"o}dinger$-$like differential equation solution}\label{sec:infelda}
The following linear combination coefficients are derived for the transformation between The Eq. (\ref{eq:21}) and Eq. (\ref{eq:17}) that represent the $R_{n^{*}}^{n^{*}-1}$ and $R_{n^{*}}^{l^{*}}$ for $n^{*}\geq 1$,
\begin{multline}\label{eq:27}
a^{l^{*}}_{n^{*}n'^{*}}
=\left(-1\right)^{n'^{*}-l^{*}-1}
\Bigg[F_{n'^{*}+l^{*}+1}\left(n^{*}+l^{*}+1\right)\Bigg.
\\
\times \Bigg.
F_{n'^{*}-l^{*}-1}\left(n^{*}-l^{*}-1\right)
F_{n'^{*}-l^{*}-1}\left(2n'^{*}\right)
\Bigg]^{1/2},
\end{multline}
and, conversely,
\begin{multline}\label{eq:28}
\tilde{a}^{l^{*}}_{n^{*}n'^{*}}=
\left(-1\right)^{n'^{*}-l^{*}-1}
\Bigg[\dfrac{1}{F_{n^{*}-l^{*}-1}\left(2n^{*}\right)} \Bigg. 
\\
\times F_{n'^{*}+l^{*}+1}\left(n^{*}+l^{*}+1\right)
\\
\Bigg.
\times F_{n'^{*}-l^{*}-1}\left(n^{*}-l^{*}-1\right)
\Bigg]^{1/2}.
\end{multline}
The quantity $F_{s}\left(n\right)=\dfrac{\Gamma[n+1]}{\Gamma[s+1] \Gamma[n-s+1]}$ is the binomial coefficient with the $\Gamma\left[x\right]$ are the gamma functions. These transformation coefficients when $\epsilon_{1}\rightarrow 1$, $\epsilon_{2}\wedge\epsilon_{3}\rightarrow 0$, are equal to the ones derived for transformations between the hydrogen$-$like orbitals and the Slater$-$type orbitals with integer principal quantum numbers.\\
The first three variants in the Table \ref{table:V1n1} are subsets of the \nth{4} variant. There is a sequence that is produced from the \nth{4} variant where principal quantum number $n^{*}$ with $n^{*}>0$ and the angular momentum quantum number $l^{*}$ with $l^{*}=l, l=0,1,2,...$ . $\epsilon_{1}\rightarrow 1$, $\epsilon_{2}\rightarrow 0$, $\epsilon_{3} \neq 0$, thus, $n^{*} \geq l+\epsilon_{3}$ and $n^{*}-l^{*}-\epsilon_{3} \in \mathbb{N}_{0}$ maintains. We mean for instance, a sequence of quantum numbers as,
\begin{align*}
\begin{array}{ll}
n^{*}=0.1, 1.1, 2.1,3.1...
\\ 
l^{*}=\hspace{1mm}0,\hspace{2.8mm}1, \hspace{2.8mm}2, \hspace{2.8mm}3, ...
\end{array}
\end{align*}
Taking into accunt the orthogonality relationship for the "generalized" Laguerre polynomials,
\begin{multline}\label{eq:29}
\int_{0}^{\infty} \left(2\zeta r\right)^{p^{*}}e^{-2\zeta r}\mathcal{L}^{p^{*}}_{q^{*}-p^{*}}\left(2\zeta r\right)
\mathcal{L}^{p^{*}}_{{q'}^{*}-{p'}^{*}}\left(2\zeta r\right)dr
\\
=\dfrac{\Gamma\left(q^{*}+1\right)}{\left(2\zeta\right)^{3}\Gamma\left(q^{*}-p^{*}+1\right)}\delta_{q^{*}-p^{*}, {q'}^{*}-{p'}^{*}},
\end{multline}
the complete and orthonormal function represents this sequence has the form that,
\begin{multline}\label{eq:30}
R_{n^{*}l^{*}}^{\epsilon_{3}}\left(\zeta, r\right)\\
=N_{n^{*}l^{*}}^{\epsilon_{3}}\left(\zeta\right)
(2\zeta r)^{l^{*}+\epsilon_{3}-1}e^{-\zeta r}
\mathcal{L}_{n^{*}-l^{*}-\epsilon_{3}}^{2l^{*}+2\epsilon_{3}}\left(2\zeta r\right),
\end{multline}
where,
\begin{align}\label{eq:31}
N_{n^{*}l^{*}}^{\epsilon_{3}}\left(\zeta\right)
=\Bigg[
\dfrac{\left(2\zeta\right)^{3}\Gamma \left( n^{*}-l^{*}-\epsilon_{3}+1 \right)}{\Gamma\left( n^{*}+l^{*}+\epsilon_{3}+1\right)}
\Bigg]^{1/2}.
\end{align}
Simplification of Laguerre polynomials in the Eq. (\ref{eq:31}) by keeping the term of highest power of $r$ reduces to the Slater$-$type orbitals with noninteger principal quantum numbers given in the Eq. (\ref{eq:21}). If $\epsilon_{3}=1$, then the Eq. (\ref{eq:31}) also reduces to the representation of the Eq. (\ref{eq:17}) in the $L^{2}\left(\mathbb{R}^{3} \right)$.
\subsection{Representing the Infeld's Schr{\"o}dinger$-$like differential equation solution in the weighted Hilbert space $L_{r^{\alpha}}^{2}\left(\mathbb{R}^{3} \right)$}\label{sec:infeldb}
Following the similar route with Guseinov \cite{19_Guseinov_1980} for generalizing the orthonormal Slater$-$type orbitals $\left( \chi_{nlm} \right)$, the $R_{n^{*}l^{*}}^{\epsilon_{3}}$ in the weighted Hilbert space $L_{r^{\alpha}}^{2}\left(\mathbb{R}^{3} \right)$ are defined as, 
\begin{multline}\label{eq:32}
R_{n^{*}l^{*}}^{\alpha \epsilon_{3}}\left(\zeta, r\right)\\
=N_{n^{*}l^{*}}^{\alpha \epsilon_{3}}\left(\zeta\right)
(2\zeta r)^{l^{*}+\epsilon_{3}-1}e^{-\zeta r}
\mathcal{L}_{n^{*}-l^{*}-\epsilon_{3}}^{2l^{*}+2\epsilon_{3}-\alpha}\left(2\zeta r\right),
\end{multline}
with,
\begin{align}\label{eq:33}
N_{n^{*}l}^{\alpha\epsilon_{3}}\left(\zeta\right)
=\Bigg[
\dfrac{\left(2\zeta\right)^{3}\Gamma n^{*}-\epsilon_{3}-l+1]}{\left(2n^{*}\right)^{\alpha}\Gamma n^{*}+\epsilon_{3}+l+1-\alpha]}
\Bigg]^{1/2}.
\end{align}
The transformation relationships where, $\epsilon=\epsilon_{3}$ in this case are given as,
\begin{multline}\label{eq:34}
a^{\alpha\epsilon l^{*}}_{n^{*}{n'}^{*}}
=\left(-1\right)^{{n'}^{*}-l^{*}-\epsilon}
\\
\times \Bigg[
\dfrac{\Gamma[ {n'}^{*}+l^{*}+\epsilon+1]}{\left(2n^{*}\right)^{\alpha}\Gamma[ {n'}^{*}+l^{*}+\epsilon+1-\alpha]} \Bigg. \\
\times \Bigg.
F_{{n'}^{*}+l^{*}+\epsilon-\alpha}\left(n^{*}+l^{*}+\epsilon-\alpha\right)
\Bigg.
\\ F_{{n'}^{*}-l^{*}-\epsilon}\left(n^{*}-l^{*}-\epsilon\right)
F_{{n'}{*}-l^{*}-\epsilon}\left(2{n'}^{*}\right)
\Bigg]^{1/2},
\end{multline}
\begin{multline}\label{eq:35}
\tilde{a}^{\alpha\epsilon l^{*}}_{n^{*}{n'}^{*}}
=\left(-1\right)^{{n'}^{*}-l^{*}-\epsilon}
\\
\times \Bigg[\dfrac{\left(2{n'}^{*}\right)^\alpha \Gamma[ n^{*}+l^{*}+\epsilon+1-\alpha ]}{\Gamma[ n^{*}+l^{*}+\epsilon+1]F_{n^{*}-l^{*}-\epsilon}\left(2n^{*}\right)}
\Bigg.\\
\times F_{{n'}^{*}+l^{*}+\epsilon-\alpha}\left(n^{*}+l^{*}+\epsilon-\alpha\right) \Bigg.
\\
\times F_{{n'}^{*}-l^{*}-\epsilon}\left(n^{*}-l^{*}-\epsilon\right)
\Bigg]^{1/2}.
\end{multline}
Finally, the transformations between $n^*-$STO and the $\Psi^{\alpha \varepsilon}$ are given as,
\begin{align}\label{eq:36}
\Psi^{\alpha \epsilon}_{n^{*}l^{*}m^{*}}\left(\zeta,\vec{r}\right)
=\sum_{n'^{*}=l^{*}+\epsilon}^{n^{*}}
a^{\alpha \epsilon l^{*}}_{n^{*}{n'}^{*}}\eta_{n'^{*}l^{*}m^{*}}\left(\zeta,\vec{r}\right),
\end{align}
\begin{align}\label{eq:37}
\eta_{n^{*}l^{*}m^{*}}\left(\zeta,\vec{r}\right)
=\sum_{{n'}^{*}=l^{*}+\epsilon}^{n^{*}}
\tilde{a}^{\alpha \epsilon l^{*}}_{n^{*}{n'}^{*}}\Psi^{\alpha \epsilon}_{{n'}^{*}l^{*}m^{*}}\left(\zeta,\vec{r}\right).
\end{align}
where, $l^{*}=l, l \in \mathbb{N}_{0}$, $m^{*}=m, m \in \mathbb{Z}$, and,
\begin{align}\label{eq:38}
\Psi^{\alpha \epsilon}_{n^{*}lm}\left(\zeta, \vec{r}\right)
=R_{n^{*}l}^{\alpha \epsilon}\left(\zeta, r\right)
S_{lm}\left(\theta, \varphi\right).
\end{align}
The $\Psi^{\alpha \epsilon}_{n^{*}lm}$ satisfy the orthonormality relationship:
\begin{multline}\label{eq:39}
\int \left[ \Psi^{\alpha \epsilon}_{n^{*}lm}\left(\zeta, \vec{r}\right) \right]^{*}
\left(\dfrac{{n'}^{*}}{\zeta r}\right)^{\alpha}
\Psi^{\alpha \epsilon}_{n^{*}l'm'}\left(\zeta, \vec{r}\right)dV
\\
=\delta_{n^{*}{n'}^{*}}\delta_{l{l'}}\delta_{m{m'}}.
\end{multline}
The variationally optimum values of principal quantum numbers $n^{*}$ listed by Koga \cite{5_Koga_1997} using the $n^*-$STOs in calculation of atoms via the Hartree$-$Fock method show that the nominative representation for the quantum numbers as $n^{*}_{1s}< \left\lbrace \begin{array}{rr}
n^{*}_{2s}
\\
n^{*}_{2p}
\end{array} \right\rbrace <n^{*}_{3s}...$
do not fulfilled. Yet, any $n^*-$STO constructed using the values of quantum numbers listed in \cite{5_Koga_1997} can be expanded or translated from one$-$point to another using the complete and orthonormal function given in the Eq. (\ref{eq:38}).
\section{Further generalization. Fractional complete and orthonormal functions}\label{sec:SchRevisit}
\subsection{Revisiting the Schr{\"o}dinger equation}\label{sec:SchRevisita}
The noninteger powers are used in the fractional calculus for a long time. They have been a subject of study in the fractal geometry as re$-$scaling an anomalous dimension $\left(D, D\equiv \epsilon\right)$ \cite{40_Butera_2014}. A non$-$Newtonian generalization for derivative dealing with the measurement of a fractal was defined in \cite{41_Liang_2019}. The connection figured out in \cite{42_Chen_2010} by proving the exsistence for \textit{local fractional derivative} \cite{43_Kolwankar_1997} that obey the following relationship,
\begin{multline}\label{eq:40}
D_{x}^{\epsilon}f\left(x\right)=\dfrac{d^{\epsilon} f\left(x\right)}{dx^{\epsilon}}
\\
=\dfrac{1}{\Gamma\left(1-\epsilon\right)}
\dfrac{d}{dx}\int_{0}^{x}\dfrac{f\left(t\right)}{\left(x-t\right)^{\epsilon-1}}dt.
\end{multline}
In addition to difficulty of the right$-$hand side of the Eq. (\ref{eq:40}) in the applications, it also does not fulfill some formula that the standard derivative have \cite{44_Khalil_2020}. Derivative of a constant, derivative for product of two functions, derivative of the quotient of two functions and chain rule for instance.
A local fractional derivative $D^{\epsilon}$ is so called as \textit{conformable derivative} of a function in \cite{44_Khalil_2020} defined as,
\begin{align}\label{eq:41}
\left(D_{t}^{\epsilon}f\right)\left(t\right)
=\lim_{\epsilon'\rightarrow 0}\dfrac{f\left(t+\epsilon' t^{1-\epsilon}\right)-f\left(t\right)}{\epsilon'}.
\end{align}
This form for the fractional calculus operator has become more popular lately because of a major flaw in the Caputo$–$Fabrizio, Atangana$–$Baleanu descriptions \cite{45_Ortigueira_2019}.

In the above sections it is already proved that an orthonormal function to be used for evaluation of integrals arising molecular Hartree$-$Fock calculations using the $n^{*}-$STOs have the property that $n^{*}-l^{*}-\epsilon \in \mathbb{N}_{0}$. The discussion ongoing in this section is for seek of completeness from the mathematical point of view. The origin for existence of the Eq. (\ref{eq:25}) is presented via the fractional calculus and its applications in the quantum mechanics. The sets of values of quantum numbers for $n^{*}-l^{*}-\epsilon$ is now extended to real space. For both $\epsilon_{1}$, $\epsilon_{3}$ if $\epsilon_{1} \wedge \epsilon_{3}>0$, $\epsilon_{1}=\epsilon_{3}=\epsilon$, $\epsilon<1$ thus, $l^{*}=l\epsilon$, $n^{*}=l\epsilon+\epsilon$. From \nth{4} variant in the table \ref{table:V1n1}, we have $\left(n-l-1\right)\epsilon$. Taking into account the presumed power series solution in the Eq. (\ref{eq:26}) for a polynomial with fractional order, this sequence of quantum numbers is meaningful since it has a form that $q^{*}=q\epsilon, q\in \mathbb{N}_{0}$. \\
A conformable \textit{fractional Schr{\"o}dinger$-$like differential equation} results with this sequence has recently been presented in spherical polar coordinates as \cite{46_Chung_2020}, 
\begin{align}\label{eq:42}
\left\lbrace
\bigtriangledown^{2\epsilon}
-\dfrac{2m^{\epsilon}}{\hbar^{2\epsilon}}
\left[
V_{\epsilon}\left(r^{\epsilon}\right)-E^{\epsilon}
\right]
\right\rbrace \Psi^{\epsilon}\left(r^{\epsilon}, \theta^{\epsilon}, \varphi^{\epsilon}\right)=0.
\end{align}
The solution that is represented by production of radial and angular parts and separates the differential equation into two parts is:
\begin{align}\label{eq:43}
\Psi_{n^{*}l^{*}m^{*}}^{\epsilon}\left(r^{\epsilon}\right)
=R_{n^{*}l^{*}}^{\epsilon}\left(r^{\epsilon}\right)
S_{l^{*}m^{*}}\left(\theta^{\epsilon}, \varphi^{\epsilon}\right).
\end{align}
The angular part,
\begin{multline}\label{eq:44}
-\left\lbrace \dfrac{1}{sin\left(\theta^{\epsilon}\right)}D_{\theta}^{\epsilon}\left[
sin\left(\theta^{\epsilon}\right)D_{\theta}^{\epsilon}S_{l^{*}m^{*}}\left(\theta^{\epsilon},\varphi^{\epsilon}\right)
\right] \right.
\\
\left. 
+\dfrac{1}{sin^{2}\left(\theta^{\epsilon}\right)}D_{\theta}^{2\epsilon}S_{l^{*}m^{*}}\left(\theta^{\epsilon},\varphi^{\epsilon}\right) \right\rbrace
=l^{*}\left(l^{*}+\epsilon\right).
\end{multline}
The radial part accordingly, has the form that,
\begin{multline}\label{eq:45}
\Bigg\{
-\frac{\hbar^{2\epsilon}}{2m^{\epsilon}}
\left[
\dfrac{1}{r^{2\epsilon}}
\left(2\epsilon r^{\epsilon}D_{r}^{\epsilon}
+r^{2\epsilon}D_{r}^{2\epsilon}
\right)-\dfrac{l^{*}\left(l^{*}+\epsilon\right)}{r^{2\epsilon}}
\right]
\Bigg.
\\
\Bigg.
-V_{\epsilon}\left(r^{\epsilon}\right)
\Bigg\} R_{n^{*}l^{*}}^{\epsilon}\left(r^{\epsilon}\right)
=E^{\epsilon}R_{n^{*}l^{*}}^{\epsilon}\left(r^{\epsilon}\right)
\end{multline}
The $m^{\epsilon}$, $\hbar^{\epsilon}$ arise due to $\epsilon$ counterpart of the standard mass and Plack constant. The standard Hamiltonian is replaced by its conformable form that is referred to as $\epsilon-$Hamiltonian \cite{46_Chung_2020}. Thus, the position and the momentum operators have their $\epsilon$ dimension of length and momentum respectively.\\
The conformable potential for hydrogen$-$like atoms therefore, is read as,
\begin{align}\label{eq:46}
V_{\epsilon}\left(r^{\epsilon}\right)=-\kappa^{\epsilon}\dfrac{Z^{\epsilon}}{r^{\epsilon}},
\end{align} 
where, $\kappa=\frac{e^{2}}{4\pi\varepsilon_{0}}$, $\varepsilon_{0}$ is the vacuum permitivity. In the light of the Bohr's theory, the substitutions to make in the Eq. (\ref{eq:44}) dimensionless as,
\begin{align}\label{eq:47}
a_{0}^{\epsilon}=\dfrac{\hbar^{2\epsilon}}{m^{\epsilon}Z^{\epsilon}\kappa^{\epsilon}},
\hspace{3mm}
-\dfrac{1}{\lambda}=\left(n\right)^{2\epsilon}=-\dfrac{m^{\epsilon}\kappa^{2\epsilon}Z^{2\epsilon}}{2\hbar^{2\epsilon}E^{\epsilon}}.
\end{align}
$n^{\epsilon}=n^{*}$. The energy eigenvalue which permits to use the atomic units $\left(\hbar=1, m=1\right)$ is first obtained as,
\begin{align}\label{eq:48}
\left(\dfrac{2m E}{\hbar^{2}}\right)^{\epsilon}
=\dfrac{\lambda^{\epsilon}}{a_{0}^{2\epsilon}}
\hspace{3mm}
\dfrac{2m^{\epsilon} Z^{\epsilon}\kappa^{\epsilon}}{\hbar^{2\epsilon}}
=\dfrac{2}{a_{0}^{\epsilon}}
\end{align}
By substituting Eqs. (\ref{eq:47}, \ref{eq:48}) into the Eq. (\ref{eq:45}) we have,
\begin{multline}\label{eq:49}
\Bigg[
D_{r}^{2\epsilon}+\dfrac{2\epsilon}{r^{\epsilon}}D_{r}^{\epsilon}
+\dfrac{\lambda}{a_{0}^{2\epsilon}}
-\dfrac{l^{*}\left(l^{*}+1\right)}{r^{2\epsilon}}
\Bigg]R^{\epsilon}\left(r^{\epsilon}\right)\\
=-\dfrac{2}{a_{0}^{\epsilon}}\dfrac{1}{r^{\epsilon}}R^{\epsilon}\left(r^{\epsilon}\right)
\end{multline}
Further simplifications is possible since the conformable derivative has the property that,
\begin{align}\label{eq:50}
D^{\epsilon}\left(fg\right)
=fD^{\epsilon}\left(g\right)
+gD^{\epsilon}\left(f\right).
\end{align}
Changing the variable from $R^{\epsilon}\left(r^{\epsilon}\right)$ to $f^{\epsilon}\left(r^{\epsilon}\right)$ as,
\begin{align}\label{eq:51}
R^{\epsilon}\left(r^{\epsilon}\right)=\dfrac{f^{\epsilon}\left(r^{\epsilon}\right)}{r^{\epsilon}}
\end{align}
results with a differential equation that is similar but comformable form to the standard Schr{\"o}dinger equation given in the Eq. (\ref{eq:16});
\begin{align}\label{eq:52}
\Bigg[
D_{r}^{2\epsilon}+\dfrac{2}{a_{0}^{\epsilon} r^{\epsilon}}
+\dfrac{\lambda}{a_{0}^{2\epsilon}}
-\dfrac{l^{*}\left(l^{*}+1\right)}{r^{2\epsilon}}
\Bigg]f^{\epsilon}\left(r^{\epsilon}\right)
=0.
\end{align}
Two regimes obtained from limits cases $r\rightarrow 0, r\rightarrow \infty$, are used to factor the full solution as,
\begin{align}\label{eq:53}
f^{\epsilon}\left(r^{\epsilon}\right)
=Ar^{l^{*}+\epsilon}e^{-\left(\zeta r\right)^{\epsilon}}g^{\epsilon}\left(r^{\epsilon}\right),
\end{align}
where, $\zeta=\sqrt{\dfrac{\lambda}{a_{0}^{2}}}$. Inserting the Eq. (\ref{eq:53}) into the differential equation we finally have the solution in terms of $\epsilon-$Whittaker conformable function as,
\begin{align}\label{eq:54}
f^{\epsilon}\left(\zeta, r^{\epsilon}\right)
=A W^{\epsilon}\left(-\frac{i}{\sqrt{\lambda}}, \frac{2l+1}{2}, 2i\zeta\frac{r^{\epsilon}}{\epsilon}\right).
\end{align}
$A$ is the normalization constant. The transformation between conformable Laguerre and Whittaker functions are given as,
\begin{multline}\label{eq:55}
\mathcal{L}^{p^{*}}_{q^{*}-p^{*}}\left(\frac{x^\epsilon}{\epsilon}\right)
=\left(-1\right)^{q-p}\dfrac{1}{\left(q-p\right)!} x^{-\frac{p^{*}+\epsilon}{2}}Exp\left[\dfrac{x^{\epsilon}}{2\epsilon}\right]
\\
\times W^{\epsilon}\left(\frac{p^{*}}{2}+\left(q^{*}-p^{*}\right)+\frac{\epsilon}{2},\frac{p}{2},\frac{x^\epsilon}{\epsilon}\right).
\end{multline}
Changing the variables as $\zeta=\dfrac{1}{n^{\epsilon}a_{0}^{\epsilon}}=\dfrac{1}{\overline{n}^{\epsilon}}$, $\overline{n}^{\epsilon}=\overline{n}\epsilon$, the wavefunction $R_{nl}^{\epsilon}\left(r^{\epsilon}\right)$ is written as,
\begin{multline}\label{eq:56}
R_{nl}^{\epsilon}\left(r^{\epsilon}\right)=
A\left[\dfrac{2}{\epsilon^{2}\overline{n}}r^{\epsilon}\right]^{l}
Exp\left[{-\frac{1}{\epsilon^{2}\overline{n}}r^{\epsilon}}\right]
\\
\times \mathcal{L}^{2l+1}_{\epsilon\left(\overline{n}-l-1\right)}\left(\dfrac{2}{\epsilon^{2}\overline{n}}r^{\epsilon}\right).
\end{multline}
\subsection{On the Laguerre functions}\label{sec:SchRevisita}
There are some studies that have been performed for fractional Laguerre functions \cite{47_Ahmed_1999, 48_Ahmed_2000, 49_Ahmed_2004, 50_Mirevski_2010, 51_Miana_2021} and an alternative solution of hydrogen atom differential equation \cite{52_Bildstein_2018, 53_Bildstein_2018}. They however avoid the local geometric behaviors \cite{42_Chen_2010, 46_Chung_2020, 54_Jeng_2010}. The derivatives directly considered only via the right$-$hand side of the Eq. (\ref{eq:40}). In some of these studies \cite{52_Bildstein_2018, 53_Bildstein_2018} the formulas derived as a generalization of the Schr{\"o}dinger equation to fractional order whereas they are for sets of quantum numbers already have representation by the first three columns of the Table \ref{table:V1n1}. We have shown in above sections that the stepwise power series expansion is available for such generalization thus, the \textit{standart} quantum mechanical formalism can be implemented. 

The set of quantum numbers defined in the Section \ref{sec:SchRevisit} indicates a transition between the first three and \nth{4} columns of the Table \ref{table:V1n1}. Such transitional set of quantum numbers are also reveals the necessity for existence of transitional Laguerre functions. The usual calculus operators ($\epsilon=1$ in the Eq. (\ref{eq:41}) are sufficient to use in generalized Laguerre functions arising in the Eq. (\ref{eq:30}) while their simpler forms require the fractional calculus operators. The resulting power series expansion for a function via the conformable fractional derivative operators has the form that,
\begin{align}\label{eq:57}
y=\sum_{r=0}^{\infty} a_{r}x^{\left(r+s\right)\epsilon}.
\end{align}
The Eq. (\ref{eq:57}) as a special case of the Eq. (\ref{eq:26}) does not fully represents the local geometric behavior for fractional derivative. The generalized Laguerre functions in the Eq. (\ref{eq:30}) can not be expressed by conformable differential equation given as \cite{55_Dixit_2020},
\begin{multline}\label{eq:58}
x^{\epsilon}\dfrac{d^{2\epsilon}y\left(x^{\epsilon}\right)}{dx^{2\epsilon}}+ \left(p\epsilon+\epsilon-x^{\epsilon}\right)\dfrac{d^{\epsilon}y\left(x^{\epsilon}\right)}{dx^{\epsilon}}\\
+\epsilon\left(q-p\right)y\left(x^{\epsilon}\right)=0.
\end{multline} 
This is because in the Eq. (\ref{eq:30}), $\left(q^{*}-p^{*}\right)=\left(q-p\right)$ and $\epsilon=1$ while for its other components $\left\lbrace q^{*}, p^{*}\right\rbrace \in \mathbb{R}$, $\epsilon \neq 1$. The above equation is obtained from Rodriguez formula [the Eq. (\ref{eq:25}) or later in this section] by replacing the variable $x$ with $\dfrac{x^{\epsilon}}{\epsilon}$ in order to make it conformable for a transformation such as,
\begin{align}\label{eq:59}
\int_{0}^{x}f\left(t\right)d^{\epsilon}t=\int_{0}^{x}f\left(t\right)\left(1-t\right)^{\epsilon-1}dt,
\end{align}
where, the following relationship between conformable and standard derivative is used \cite{56_Atangana_2015};
\begin{align}\label{eq:60}
D^{n\epsilon}f\left(x\right)=x^{n-n\epsilon}\dfrac{f^{\left(n\right)}\left(x\right)}{n!}.
\end{align}
A power series solution for the Eq. (\ref{eq:58}) was made in \cite{57_Shat_2019}. The problem with the conformable local derivative is that the power series expansion requires a clear separation between integer and fractional parts as $q \epsilon$. As it is stated above the parameters do not always meet this condition.

Since they are special case of the generalized Laguerre functions with integer order, the transitional Laguerre functions with noninteger order offer an excellent opportunity to numerically test the accuracy of a fractional calculus method. Investigation of a special function through the fractional calculus operators seperates into two parts. Generalization of parameters to fractional order and generalization of corresponding differential equation to fractional form. Power series representation is helpful to characterize of these two concepts. A power series expansion that covers the both is given in the Eq. (\ref{eq:26}). In this expansion $\epsilon=1$ corresponds to generalization of parameters while $\epsilon'=0$ corresponds to generalization of differential equation of the special function. The transitional Laguerre polynomials comprise the both cases and they form according to decision to be made. The discussion is made for $\epsilon'=0$ in the previous section. A comprehensive study made in \cite{49_Ahmed_2004} for the Rodrigues representation \cite{58_Rasala_1981, 59_Weber_2007} of Laguerre functions with noninteger orders. The Riemann-Liouville type fractional calculus operators were considered. The generalized Laguerre polynomials after a slightly modification is written as,
\begin{multline}\label{eq:61}
\mathcal{L}^{p^{*}}_{q^{*}-p^{*}}\left(x\right)
\\
=\dfrac{{e^{x}}x^{-p^{*}}}{\Gamma\left(q^{*}-p^{*}+1\right)}
\mathcal{Y}^{\hspace{0.5mm} p^{*}}_{q^{*}-p^{*}}\left(q^{*}-p^{*},1,x\right),
\end{multline}
where,
\begin{align}\label{eq:62}
\mathcal{Y}^{b}_{a}\left(c,d,x\right)
=D^{a}_{x}x^{b+c}e^{-cx}.
\end{align}
The transitional Lagurre functions are accordingly expressed as,
\begin{align}\label{eq:63}
\mathcal{L}^{0}_{q^{*}}\left(x\right)
=\dfrac{{e^{x}}}{\Gamma\left(q^{*}+1\right)}
\mathcal{Y}^{\hspace{0.5mm} p^{*}}_{q^{*}}\left(q^{*}-p^{*},1,x\right).
\end{align}
The Eq. (\ref{eq:61}) has the following series representation \cite{38_Magnus_1966},
\begin{multline}\label{eq:64}
\mathcal{L}^{p^{*}}_{q^{*}-p^{*}}\left(x\right)
=\sum_{k=0}^{\infty}\dfrac{\Gamma\left(q^{*}+1\right)}{\Gamma\left(q^{*}-p^{*}-k+1\right)\Gamma\left(p^{*}+k+1\right)}
\\
\times \dfrac{\left(-x\right)^{k}}{\Gamma\left(k+1\right)}.
\end{multline}
Using the Leibniz rule \cite{60_Osler_1971}, for $\epsilon$, $\epsilon \in \left(n,n-1\right)$, $n \geq 1$,
\begin{align}\label{eq:65}
D_{x}^{\epsilon}\left[f\left(x\right)g\left(x\right)\right]
=\sum_{k=0}^{\infty}
F_{k}\left(\epsilon\right)
D^{k}_{x}f\left(x\right)D^{\epsilon-k}_{x}g\left(x\right),
\end{align}
we may have,
\begin{align}\label{eq:66}
\mathcal{L}^{p^{*}}_{q^{*}-p^{*}}\left(x\right)
=\left(-1\right)^{q^{*}-p^{*}}D^{p^{*}}_{x}\mathcal{L}^{0}_{q^{*}}\left(x\right).
\end{align}
Although the right$-$hand side of the Eq. (\ref{eq:66}) is fractional derivative of a transitional Laguerre function, the left$-$hand side can be a generalized Laguerre function with integer order $q^{*}-p^{*}$ even though $\left\lbrace q^{*},p^{*}\right\rbrace \in \mathbb{R}^{+}$. Thus the accuracy of the fractional calculus operators and the relationships derived based on them for the Laguerre functions are now accessible to verify in both conformable or Riemann$-$Liouville type fractional derivative operators. The Laguerre functions are also complete in the weighted Hilbert space $L_{w^{\left(p^{*},\epsilon\right)}}$ according to inner product given as \cite{61_Bhrawy_2016},
\begin{align}\label{eq:67}
\left(f\vert g\right)_{w^{\left(p^{*},\epsilon\right)}}
\int_{-\infty}^{\infty}w^{\left(p^{*},\epsilon\right)}\left(x\right)f^{*}\left(z\right)g\left(z\right)dz.
\end{align} 
If the Laguerre functions are studied in the conformable form then the transformation from anomalous dimension let say $d^{\epsilon}t$ to $dt$ should also be taken into accont [Eq. (\ref{eq:59})].
\subsection{Fractional complete and orthonormal functions in the weighted Hilbert space $L_{r^{\alpha\epsilon}}^{2}\left(\mathbb{R}^{3\epsilon} \right)$}\label{sec:SchRevisitc}
A Sturmian$-$like wavefunction closely related to hydrogen eigen$-
$function given in the Eq. (\ref{eq:56}) is written as,
\begin{multline}\label{eq:68}
R_{nl}^{\epsilon}\left(\zeta, r^{\epsilon}\right)=
N^{\epsilon}_{n,l}\left(\zeta\right)\left(2\zeta \dfrac{r^{\epsilon}}{\epsilon}\right)^{l}
Exp\left[-\zeta \frac{r^{\epsilon}}{\epsilon}\right]
\\
\times \mathcal{L}^{2l+1}_{\epsilon\left(n-l-1\right)}\left(2\zeta \dfrac{r^{\epsilon}}{\epsilon}\right).
\end{multline}
The following orthogonality relationship is obtained by modification of the Eq. (\ref{eq:29}) according to the Eq. (\ref{eq:68}) that is representation of the wavefunction in the conformable $\epsilon-$dimension,
\begin{multline}\label{eq:69}
\int_{0}^{\infty} Exp\left[-2\zeta \dfrac{r^{\epsilon}}{\epsilon}\right]\left(2\zeta r^{\epsilon}\right)^{p}
\\
\times \mathcal{L}^{p}_{\epsilon\left(q-p\right)}\left(2\zeta \dfrac{r^{\epsilon}}{\epsilon}\right)
\mathcal{L}^{p}_{\epsilon\left(q'-p'\right)}\left(2\zeta \dfrac{r^{\epsilon}}{\epsilon}\right)d^{\epsilon}r
\\
=\dfrac{\epsilon^{l}\Gamma\left(q+1\right)}{\left(2\zeta\right)^{3}\Gamma\left(q-p+1\right)}\delta_{q-p, {q'}-{p'}}.
\end{multline}
$R_{nl}^{\epsilon}\left(\zeta, r^{\epsilon}\right)$ are derived from the general Sturm$-$Liouville eigenvalue problem \cite{62_Wang_2021} they are thus, complete and orthonormal in fractional Hilbert space, $L_{r^{\epsilon}}^{2}\left(\mathbb{R}^{3\epsilon} \right)$ with respect to $\dfrac{n\epsilon}{\zeta r^{\epsilon}}$:
\begin{align}\label{eq:70}
\int_{0}^{\infty}
R_{nl}^{\epsilon}\left(\zeta, r^{\epsilon}\right)
\left(\dfrac{n'\epsilon}{\zeta r^{\epsilon}}\right)
R_{n'l}^{\epsilon}\left(\zeta, r^{\epsilon}\right)
r^{2\epsilon}d^{\epsilon}r
=\delta_{n,n'}.
\end{align}
The wavefunction given in the Eq. (\ref{eq:68}) is also available to be written in a weighted $L_{r^{\alpha\epsilon}}^{2}\left(\mathbb{R}^{3\epsilon} \right)$ space with respec to $\left(\dfrac{n\epsilon}{\zeta r^{\epsilon}}\right)^{\alpha}$ as follows,
\begin{multline}\label{eq:71}
R_{nl}^{\alpha\epsilon}\left(\zeta, r^{\epsilon}\right)=
N^{\alpha\epsilon}_{nl}\left(\zeta\right)\left(2\zeta \dfrac{r^{\epsilon}}{\epsilon}\right)^{l}
Exp\left[-\zeta \frac{r^{\epsilon}}{\epsilon}\right]
\\
\times \mathcal{L}^{2l+2-\alpha}_{\epsilon\left(n-l-1\right)}\left(2\zeta \dfrac{r^{\epsilon}}{\epsilon}\right),
\end{multline}
where, the normalization constant is obtained as,
\begin{align}\label{eq:72}
N^{\alpha\epsilon}_{nl}\left(\zeta\right)
=\dfrac{1}{\epsilon^{l+1}}
\Bigg[
\dfrac{\left(2\zeta\right)^{3}\Gamma n-l]}{\left(2n\right)^{\alpha}\Gamma n+l+2-\alpha]}
\Bigg]^{1/2}.
\end{align} 
\section{Relationship between Riemann$-$Liouville type integrals and relativistic molecular auxiliary functions}\label{sec:rmmolaux}
\begin{figure*}[htp!]
    \centering
    \includegraphics[width=1.0\textwidth]{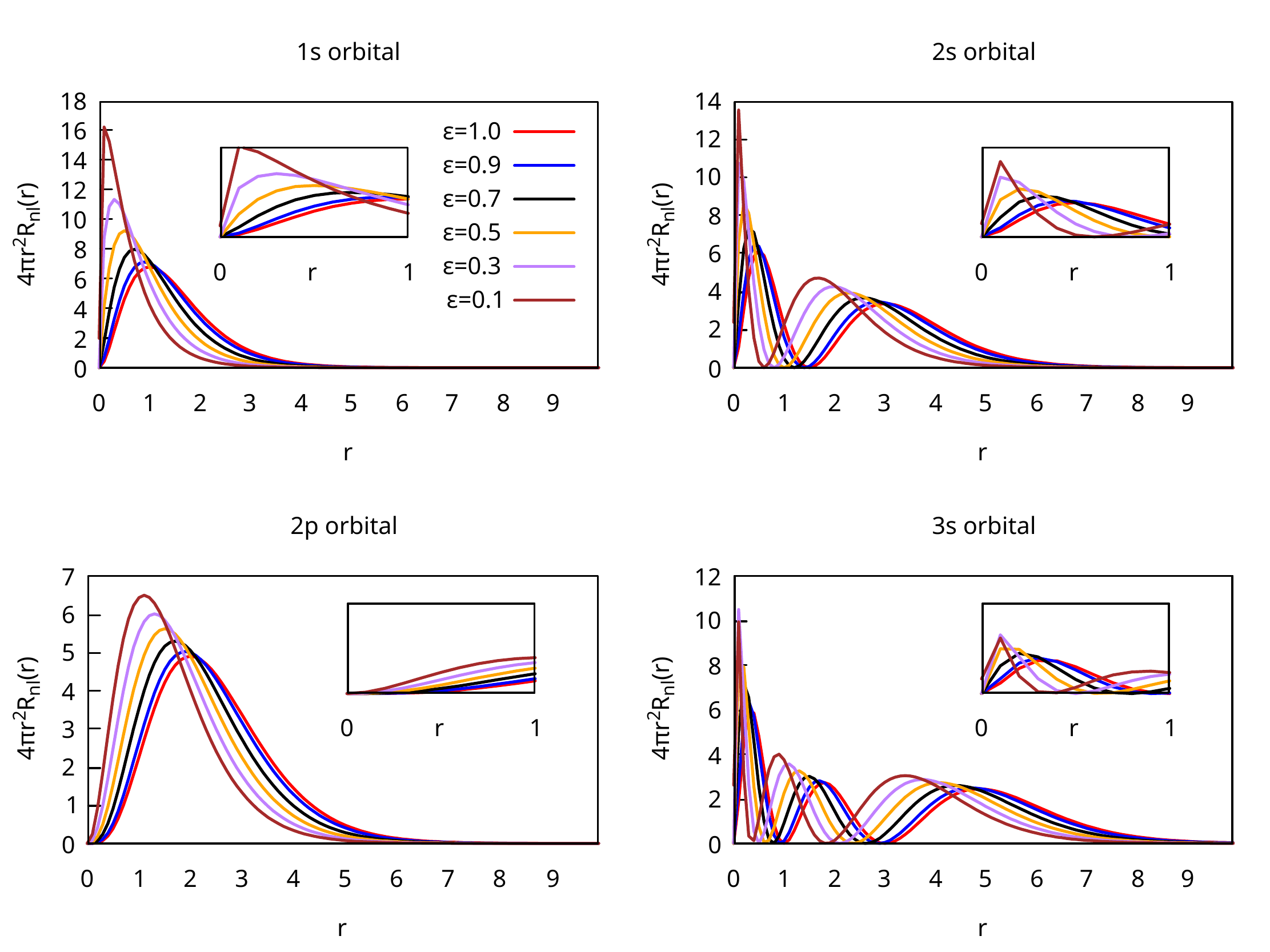}
    \caption{Radial distribution functions $\left\lbrace 4\pi r^{2}\left[R^{\epsilon}_{n^{*}l^{*}}\left(\zeta, r\right)\right]^{*} R^{\epsilon}_{n^{*}l^{*}}\left(\zeta, r\right) \right\rbrace$ for different values of $\epsilon$ and in atomic units $\left(a.u.\right)$. They are obtanied through the Eq. (\ref{eq:30}) where $\zeta=1$. $\epsilon=1$ coresponds to Lambda functions.}
    \label{fig:infeld2}
\end{figure*}
The auxiliary functions given in the Eq. (\ref{eq:13}) by changing the variable $\mu$ as,
\begin{align}\label{eq:73}
\xi=\frac{1}{1-\xi'}\Rightarrow
d\xi=\frac{1}{\left(1-\xi'\right)^{2}}d\xi',
\end{align}
have the below form,
\begin{multline} \label{eq:74}
\left\lbrace \begin{array}{cc}
\mathcal{P}^{n_1,q}_{n_{2}n_{3}n_{4}}\left(p_{123} \right)
\\
\mathcal{Q}^{n_1,q}_{n_{2}n_{3}n_{4}}\left(p_{123} \right)
\end{array} \right\rbrace
=\frac{p_{1}^{\sl n_{1}}}{\left({\sl n_{4}}-{\sl n_{1}} \right)_{\sl n_{1}}}
\\
\times \int_{0}^{1}\int_{-1}^{1}\nu^{q}\left(1+\nu-\xi'\nu \right)^{n_{2}}\left(1-\nu+\xi'\nu \right)^{n_{3}}
\\
\times \left(1-\xi'\right)^{-q-n_{2}-n_{3}-2}
\left\lbrace \begin{array}{cc}
P\left[{\sl n_{4}-n_{1}},p_{1}\mathfrak{f}_{ij}^{k}\left(\frac{1}{1-\xi'}, \nu\right) \right]
\\
Q\left[{\sl n_{4}-n_{1}},p_{1}\mathfrak{f}_{ij}^{k}\left(\frac{1}{1-\xi'}, \nu\right) \right]
\end{array} \right\rbrace
\\
\times
e^{p_{2}\left(\frac{1}{1-\xi'}\right)-p_{3}\nu}d\xi' d\nu.
\end{multline}
Note that, the domain for the integrals given in this form of molecular auxiliary functions is more advantageous in terms of numerical calculations. The Riemann$-$Liouville type integrals may be derived via the Eq. (\ref{eq:74}). Similarly to the Eq. (\ref{eq:73}) we change the $\xi$ variable this time with,
\begin{align}\label{eq:75}
\xi=\frac{\tau}{\tau-\xi'}\Rightarrow
d\xi=\frac{\tau}{\left(\tau-\xi'\right)^{2}}d\xi',
\end{align}
thus,
\begin{multline} \label{eq:76}
\left\lbrace \begin{array}{cc}
\mathcal{P}^{n_1,q}_{n_{2}n_{3}n_{4}}\left(p_{123} \right)
\\
\mathcal{Q}^{n_1,q}_{n_{2}n_{3}n_{4}}\left(p_{123} \right)
\end{array} \right\rbrace
=\frac{p_{1}^{\sl n_{1}}}{\left({\sl n_{4}}-{\sl n_{1}} \right)_{\sl n_{1}}}
\\
\times \int_{0}^{\tau}\int_{-1}^{1}\left(\tau\nu\right)^{q}\left(\tau+\tau\nu-\xi'\nu \right)^{n_{2}}\left(\tau-\tau\nu+\xi'\nu \right)^{n_{3}}
\\
\times \left(\tau-\xi'\right)^{-q-n_{2}-n_{3}-2}
\left\lbrace \begin{array}{cc}
P\left[{\sl n_{4}-n_{1}},p_{1}\mathfrak{f}_{ij}^{k}\left(\frac{\tau}{\tau-\xi'}, \nu\right) \right]
\\
Q\left[{\sl n_{4}-n_{1}},p_{1}\mathfrak{f}_{ij}^{k}\left(\frac{\tau}{\tau-\xi'}, \nu\right) \right]
\end{array} \right\rbrace
\\
\times
e^{p_{2}\left(\frac{\tau}{\tau-\xi'}\right)-p_{3}\nu}d\xi' d\nu.
\end{multline}
Several kernels for the fractional calculus can be derived from the Eq. (\ref{eq:76}). Here, we focus on the simplest Riemann$-$Liouville type kernel and show that it is a special case of the Eq. (\ref{eq:13}). By representing the integral over $\nu$ as a function and embedding the coefficient $\frac{p_{1}^{n_{1}}}{\left(n_{4}-n1\right)_{n_{1}}}$ into this function, product$/$divide the expression with $\Gamma\left[ -q-n_{2}-n_{3}-1\right]$, the Eq. (\ref{eq:76}) takes the form that,
\begin{multline}\label{eq:77}
\left\lbrace \begin{array}{cc}
\mathcal{P}^{n_1,q}_{n_{2}n_{3}n_{4}}\left(\tau,p_{123} \right)
\\
\mathcal{Q}^{n_1,q}_{n_{2}n_{3}n_{4}}\left(\tau, p_{123} \right)
\end{array} \right\rbrace
=\frac{1}{\Gamma\left[q-n_{2}-n_{3}-1\right]}\\
\times
\int_{0}^{\tau}
\left(\tau-\xi'\right)^{-q-n_{2}-n_{3}-2}
\\
\times \left\lbrace \begin{array}{cc}
^{1}f^{n_{1}}_{n_{2}n_{3}n_{4}}\left(\tau, \xi',\nu,p_{123}\right)
\\
^{2}f^{n_{1}}_{n_{2}n_{3}n_{4}}\left(\tau, \xi',\nu,p_{123}\right)
\end{array} \right\rbrace
d\xi'.
\end{multline}
The well$-$known Riemann$-$Liouville integral is analogous to the Eq. (\ref{eq:77}) with $\beta=-q-n_{2}-n_{3}-1$, $\beta \in \mathcal{R}$ for any analytical function that can be represented by the integrals over $\nu$. The $^{i}f^{n_{1}}_{n_{2}n_{3}n_{4}}\left(\tau, \xi',\nu,p_{123}\right)$, $i=1,2$ are arbitrary functions. For these functions it is not necessarily true that they depend on $\tau$. Such possible dependency however, forms more generic integrals which we previously have referred to as relativistic auxiliary functions integrals \cite{11_Bagci_2015}. In the light of this fact we can write,
\begin{multline}\label{eq:78}
\left\lbrace \begin{array}{cc}
\mathcal{P}^{n_1,q}_{n_{2}n_{3}n_{4}}\left(\tau,p_{123} \right)
\\
\mathcal{Q}^{n_1,q}_{n_{2}n_{3}n_{4}}\left(\tau, p_{123} \right)
\end{array} \right\rbrace
\\
=\left\lbrace \begin{array}{cc}
I^{\beta}\left[^{1}f^{n_{1},\beta}_{00 n_{n4}}\left(\xi',p_{123} \right)\right]
\\
I^{\beta}\left[^{2}f^{n_{1},\beta}_{00 n_{n4}}\left(\xi',p_{123} \right)\right]
\end{array} \right\rbrace
\\
=\frac{1}{\Gamma\left[-\beta\right]}
\int_{0}^{\tau} \left(\tau-\xi'\right)^{-\beta-1}
\\
\times \left\lbrace \begin{array}{cc}
^{1}f^{n_{1}}_{n_{2}n_{3}n_{4}}\left(\xi',p_{123}\right)
\\
^{2}f^{n_{1}}_{n_{2}n_{3}n_{4}}\left(\xi',p_{123}\right)
\end{array} \right\rbrace
d\xi' .
\end{multline}
\section{Discussions}\label{sec:discuss}
The sets of complete and orthonormal functions to be used in atomic and molecular physics are presented for both integer and fractional orders. They are constructed according to possible variants for electron configurations given in the Table \ref{table:V1n1}. It is shown that a formula derived local derivative \cite{44_Khalil_2020} in the Eq. (\ref{eq:41}) does not holds for all variants of quantum numbers. A criterion for accuracy of a fractional calculus operator considering the Laguerre polynomials is defined in the Eq. (\ref{eq:66}). Additionally to general properties for fractional derivative of a function discussed in the Section \ref{sec:SchRevisita}, we can say that for a fractional derivative operator let say $D^{\epsilon}$ the limit $\epsilon \rightarrow 1$ and $D^{\epsilon}=D$ is necessarily bot not sufficient condition. The differential equation given in the Eq. (\ref{eq:58}) does not represents the general case. It is possible for $p^{*}$, $p\epsilon$ to be a real number $\left(\epsilon \neq 1\right)$ while $q^{*}-p^{*}=q\epsilon-p\epsilon$ is a natural number $\left(\epsilon=1\right)$. In this case Riemann$-$Liouville type fractional derivative operators [left$-$hand side of the Eq. (\ref{eq:40})] may be used. It is shown in the Section \ref{sec:rmmolaux} that they are special case of relativistic auxiliary functions.

Results for radial distribution functions of exponential$-$type functions with noninteger order are plotted in the Figures \ref{fig:infeld1} and \ref{fig:infeld2}. These figures show that the limit cases $\epsilon\rightarrow 0$, $\epsilon\rightarrow 1$ are reduce to the $R_{nl}^{\alpha}$ \cite{18_Guseinov_2002} for $\alpha=1$ and $\alpha=0$. These values of $\alpha$, $\alpha=\left\lbrace 1,0 \right\rbrace$ are represents the Coulomb$-$Sturmian and Lambda functions, respectively. Based on the sets of exponential$-$type of orthonormal functions, four variants for the Slater$-$type orbitals may be obtained by simplification of Laguerre polynomials to highest power of $r$. These orbitals are written as,
\begin{align*}
&\eta_{n}^{n-1}\left(\zeta, r\right) \hspace{3mm} [n \in \mathbb{N}_{0}, n\geq l+1]
\\
&\eta_{n^{*}}^{n^{*}-1}\left(\zeta, r\right) \hspace{1.5mm} [n^{*} \in \mathbb{R}^{+}, n^{*}\geq l+\epsilon+1]
\\
&\eta_{n^{*}}^{n^{*}-1}\left(\zeta, r\right) \hspace{1.5mm} [n^{*} \in \mathbb{R}^{+} \and n^{*}>0,  n^{*}\geq l+\epsilon]
\\
&\eta_{n^{*}}^{n^{*}-\epsilon}\left(\zeta, r\right) \hspace{1.5mm} [n^{*} \in \mathbb{R}^{+} \and n^{*}>0,  n^{*}\geq \epsilon\left(l+1\right)]
\end{align*}

For the translation of $\eta_{p}$ in terms of the orthonormal $\chi_{p}$ at a displaced center the following formulas are derived which, can be used in both analytical and numerical calculations;
\begin{multline}\label{eq:79}
\eta_{n^{*}_{p}l^{*}_{p}m^{*}_{p}}\left(\zeta, \vec{r_{a}}\right)\\
=\sum_{n'^{*}=n_{1}^{*}}^{n_{_{N}}}
\sum_{{l'}^{*}=l_{1}^{*}}^{l_{p}}
\sum_{{m'}^{*}=m_{1}^{*}}^{m_{p}}
A_{n^{*}_{p}l^{*}_{p}m^{*}_{p},n'^{*}{l'}^{*}{m'}^{*}}\left(\zeta,\zeta', \vec{R}\right)\\
\times \chi_{n'^{*}{l'}^{*}{m'}^{*}}\left(\zeta',\vec{r}_{b}\right),
\end{multline} 
where,
\begin{multline}\label{eq:80}
A_{n^{*}_{p}l^{*}_{p}m^{*}_{p},n'^{*}{l'}^{*}{m'}^{*}}\left(\zeta,\zeta', \vec{R}\right)\\
=\int \eta^{*}_{n^{*}_{p}l^{*}_{p}m^{*}_{p}}\left(\zeta, \vec{r}_{a}\right)
\chi_{n'^{*}{l'}^{*}{m'}^{*}}\left(\zeta', \vec{r}_{b}\right)dV,
\end{multline}
and,
\begin{multline}\label{eq:81}
A_{n^{*}_{p}l^{*}_{p}m^{*}_{p},n'^{*}{l'}^{*}{m'}^{*}}\left(\zeta,\zeta', \vec{R}\right)
=\sum_{n''^{*}=n'^{*}}^{N}a_{n'^{*}n''^{*}} \\
\times
\int \eta^{*}_{n^{*}_{p}l^{*}_{p}m^{*}_{p}}\left(\zeta, \vec{r}_{a}\right)
\eta_{n''^{*}{l''}^{*}{m''}^{*}}\left(\zeta', \vec{r}_{b}\right)dV.
\end{multline}
$\vec{R}=\vec{R}_{ab}=\vec{r}_{a}-\vec{r}_{b}$ is the inter$-$nuclear distance vector. The vectors $\vec{r}_{a}$, $\vec{r}_{b}$ are the radius vectors of electron with respect to nuclear labels $a$, $b$. $\left\lbrace n^{*}_{_{N}}, l_{_{N}}, m_{_{N}} \right\rbrace$ represents the principal, angular momentum and magnetic quantum numbers of the last element of a vector set, respectively. $N$ is the upper limit of summation. The two$-$center overlap integrals in the Eq. (\ref{eq:81}) are expressed in terms of molecular auxiliary functions defined in the prolate spheroidal coordinates. They can be calculated by using the methods presented in \cite{10_Bagci_2014, 15_Bagci_2020}. The computer program code \cite{36_Bagci_2022} written by one of the authors in the $Julia$ programming language is available to used for calculation of relativistic molecular auxiliary functions.

\begin{appendices}
\section*{Appendices}\label{sec:appendices}
\subsection{Gram$-$Schmidt Method}\label{sec:appendicesa}
\setcounter{equation}{0}
\renewcommand{\theequation}{A.\arabic{equation}}
The following recurrence relationships \cite{17_Werneth_2010} are used to numerically calculate the $\chi_{n^{*}lm}$ using the non$-$orthogonal but normalized $\eta_{n^{*}lm}$,
\begin{align}\label{eq:a1}
\chi_{p}=\mathfrak{N}_{p}
\Bigg[
\eta_{p}-\sum_{q=1}^{p-1}\mathfrak{w}_{qp}\chi_{p}.
\Bigg]
\end{align}
projecting $\eta_{r}$ into the Eq.(\ref{eq:15}) the $\mathfrak{w}_{qp}$ are calculated through recurrence relationships as,
\begin{align}\label{eq:a2}
\mathfrak{w}_{rp}=
\mathfrak{N}_{p}
\Bigg[
w_{rp}-\sum_{q=1}^{p-1}\mathfrak{w}_{pq}w_{rq}
\Bigg],
\end{align}
where,
\begin{align}\label{eq:a3}
\mathfrak{w}_{rp}=\int \eta_{r}^{*}\chi_{p}dV,
\end{align}
\begin{align}\label{eq:a4}
w_{rp}=\int \eta_{r}^{*}\eta_{p}dV,
\end{align}
and,
\begin{align}\label{eq:a5}
\mathfrak{N}_{p}
=\sqrt{\dfrac{1}{1-\sum_{q=1}^{p-1}\mathfrak{w}_{qp}^2}}.
\end{align}
Note that, the Eq. (\ref{eq:a2}) are useful when a large set of orthonormalized functions is needed. \\
The linear combination coefficients arising in the translation of $\chi_{n^{*}lm}$ into $\eta_{n^{*}lm}$,
\begin{align}\label{eq:a6}
\chi_{p}=\sum_{q=1}^{p} a_{pq}\eta_{q},
\end{align}
are numerically calculated as,
\begin{align}\label{eq:a7}
a_{rp}=\mathfrak{N}_{p}\sum_{q=p}^{r-1}\mathfrak{w}^{*}_{qr}a_{qp}.
\end{align}
Here, $r>p$, $\left\lbrace r,p \right\rbrace$ stand to represent the positions of elements in a vector set. The initial conditions are given as, $\mathfrak{N}_{1}=1$, $\chi_{1}=\eta_{1}$, $\mathfrak{w}_{21}=w_{21}$, $a_{pp}=1$ for any values of $p$. The inverse transformation between $\chi_{p}$ and $\eta_{p}$;
\begin{align}\label{eq:a8}
\eta_{p}=\sum_{q=1}^{p} \tilde{a}_{pq}\chi_{q},
\end{align}
\begin{align}\label{eq:a9}
\tilde{a}_{pq}=\dfrac{a_{pq}}{\mathfrak{N}_{p}}.
\end{align}
\end{appendices}


\begin{thebibliography}{plainnat}

\bibitem{1_Slater_1930} Slater JC (1930) \textit{Atomic Shielding Constants}. Physical Review A \textbf{36}(1): 57--64. doi: \url{https://link.aps.org/doi/10.1103/PhysRev.36.57}

\bibitem{2_Infeld_1951} Infeld L, Hull TE (1951) \textit{The Factorization Method}. Rev. Mod. Phys. \textbf{23}(1): 21--68:doi: \url{https://link.aps.org/doi/10.1103/RevModPhys.23.21}

\bibitem{3_Parr_1957} Parr RG, Joy HW (1957) \textit{Why Not Use Slater Orbitals of Nonintegral Principal Quantum Number?}. The Journal of Chemical Physics \textbf{26}(2): 424. doi: \url{https://doi.org/10.1063/1.1743314}

\bibitem{4_Condon_1935} Condon EU, Shortley GH (1935) \textit{The Theory of Atomic Spectra}. Cambridge University Press, Cambridge, UK

\bibitem{5_Koga_1997} Koga T, Kanayama K, Thakkar AJ (1997) \textit{Noninteger principal quantum numbers increase the efficiency of Slater-type basis sets}. International Journal of Quantum Chemistry \textbf{62}(1): 1--11. doi: \url{https://doi.org/10.1002/(SICI)1097-461X(1997)62:1<1::AID-QUA1>3.0.CO;2-\#}

\bibitem{6_Koga_1997} Koga T, Kanayama K (1997) \textit{Noninteger principal quantum numbers increase the efficiency of Slater-type basis sets: singly charged cations and anions}. Journal of Physics B: Atomic, Molecular and Optical Physics \textbf{30}(7): 1623--1631. doi: \url{https://doi.org/10.1088/0953-4075/30/7/004}

\bibitem{7_Koga_1997} Koga T, Kanayama K (1997) \textit{Noninteger principal quantum numbers increase the efficiency of Slater-type basis sets: heavy atoms}. Chemical Physics Letters \textbf{266}(1): 123--129. doi: \url{https://doi.org/10.1016/S0009-2614(96)01500-X}

\bibitem{8_Koga_1998} Koga T, Garc{\'i}a de la Vega JM, Miguel B (1998) \textit{Double-zeta Slater-type basis sets with noninteger principal quantum numbers and common exponents}. Chemical Physics Letters \textbf{283}(1): 97--101. doi: \url{https://doi.org/10.1016/S0009-2614(97)01322-5}

\bibitem{9_Koga_2000} Koga T, Shimazaki T, Satoh T (2000) \textit{Noninteger principal quantum numbers increase the efficiency of Slater-type basis sets: double-zeta approximation}. Journal of Molecular Structure: THEOCHEM \textbf{496}(1): 95--100. doi: \url{https://doi.org/10.1016/S0166-1280(99)00176-1}

\bibitem{10_Bagci_2014} Ba{\u g}c{\i} A, Hoggan PE (2014) \textit{Performance of numerical approximation on the calculation of overlap integrals with noninteger Slater-type orbitals}. Physical Review E \textbf{89}(7): 053307. doi: \url{https://doi.org/10.1103/PhysRevE.89.053307}

\bibitem{11_Bagci_2015} Ba{\u g}c{\i} A, Hoggan PE (2015) \textit{Benchmark values for molecular two-electron integrals arising from the Dirac equation}. Physical Review E \textbf{91}(2): 023303. doi: \url{https://doi.org/10.1103/PhysRevE.91.023303}

\bibitem{12_Bagci_2015} Ba{\u g}c{\i} A, Hoggan PE (2015) \textit{Benchmark values for molecular three-center integrals arising in the Dirac equation}. Physical Review E \textbf{92}(4): 043301. doi: \url{https://doi.org/10.1103/PhysRevE.92.043301}

\bibitem{13_Bagci_2018} Ba{\u g}c{\i} A, Hoggan PE (2018) \textit{Analytical evaluation of relativistic molecular integrals. I. Auxiliary functions}. Rend. Fis. Acc. Lincei \textbf{29}(1): 191--197. doi. \url{https://doi.org/10.1007/s12210-018-0669-8}

\bibitem{14_Bagci_2018} Ba{\u g}c{\i} A, Hoggan PE (2018) \textit{Analytical evaluation of relativistic molecular integrals. II: Method of computation for molecular auxiliary functions involved}. Rend. Fis. Acc. Lincei \textbf{29}(4): 765--775. doi. \url{https://doi.org/10.1007/s12210-018-0734-3}

\bibitem{15_Bagci_2020} Ba{\u g}c{\i} A, Hoggan PE (2020) \textit{Analytical evaluation of relativistic molecular integrals: III. Computation and results for molecular auxiliary functions}. Rend. Fis. Acc. Lincei \textbf{31}(4): 1089--1103. doi: \url{https://doi.org/10.1007/s12210-020-00953-3}

\bibitem{16_Allouche_1976} Allouche A (1976) \textit{Non$-$integer Slater orbital calculations}. Theoret. Chim. Acta \textbf{42}(4): 325--332. doi: \url{https://doi.org/10.1007/BF00548474}

\bibitem{17_Werneth_2010} Werneth CN, Dhar M, Maung KM, Sirola C, Norbury JW (2010) \textit{Numerical Gram-Schmidt orthonormalization}. Eur. J. Phys. \textbf{31}(3): 693--700. doi: \url{https://doi.org/10.1088/0143-0807/31/3/027}

\bibitem{18_Guseinov_2002} Guseinov II (2002) \textit{New complete orthonormal sets of exponential-type orbitals and their application to translation of Slater orbitals}. Int. J. Quant. Chem. \textbf{90}(1): 114--118. doi: \url{https://doi.org/10.1002/qua.927}

\bibitem{19_Guseinov_1980} Guseinov II (1980) \textit{Expansion of Slater-type orbitals about a new origin and analytical evaluation of multicenter electron-repulsion integrals}. Phys. Rev. A \textbf{22}(2): 369--371. doi: \url{https://link.aps.org/doi/10.1103/PhysRevA.22.369}

\bibitem{20_Polya_1976} P{\`o}lya G, Szeg{\"o} G, Billigheimer CE (1976) \textit{Problems and theorems in analysis. Vol. II}. Springer, Berlin

\bibitem{21_Hobson_1931} Hobson EW (1931) \textit{The Theory of Spherical and Ellipsoidal Harmonics.} Cambridge University Press, Cambridge, UK 

\bibitem{22_Oldham_1974} Oldham KB, Spanier J (1974) \textit{The Fractional Calculus: Theory and Applications of Differentiation and Integration to Arbitrary Order} Academic Press, New York and London

\bibitem{23_Kilbas_2006} Kilbas AA, Srivastava HM, Trujillo JJ (2006) \textit{Theory and Applications of Fractional Differential Equations} North-Holland Mathematics Studies, Elsevier, Amsterdam

\bibitem{24_Ozarslan_2010} {\"O}zarslan MA, {\"O}zergin E (2010) \textit{Some generating relations for extended hypergeometric functions via generalized fractional derivative operator}. Mathematical and Computer Modelling \textbf{52}(9): 1825--1833. doi: \url{https://doi.org/10.1016/j.mcm.2010.07.011}

\bibitem{25_Luo_2013} Luo MJ, Raina RK (2013) \textit{Extended Generalized Hypergeometric Functions and Their Applications}. Bull. Math. Anal. App. \textbf{5}(4): 65--77.

\bibitem{26_Luo_2014} Luo MJ, Milovanovic GV, Agarwal P (2014) \textit{Extended beta function, Extended Gauss hypergeometric function, Extended generalized hypergeometric function, Fractional integral, -function, Laguerre polynomial}. App. Math. and Comput. \textbf{248}: 631--651. doi: \url{https://doi.org/10.1016/j.amc.2014.09.110}

\bibitem{27_Nisar_2020} Nisar KS, Suthar DL, Agarwal R and Purohit SD (2020) \textit{Fractional calculus operators with Appell function kernels applied to Srivastava polynomials and extended Mittag-Leffler function} Adv. Differ. Equ. \textbf{148}(2020). doi: \url{https://doi.org/10.1186/s13662-020-02610-3}

\bibitem{28_Jain_2022} Jain S, Cattani C, Agarwal P (2022) \textit{Fractional Hypergeometric Functions} Symmetry \textbf{14}(4): 714. doi: \url{https://www.mdpi.com/2073-8994/14/4/714}

\bibitem{29_Gill_2012} Gill A, Segura J, Temme NM (2012) \textit{Efficient and Accurate Algorithms for the Computation and Inversion of the Incomplete Gamma Function Ratios}. Siam J. Sci. Comput. \textbf{34}(6): A2965--A2981. doi: \url{https://doi.org/10.1137/120872553}

\bibitem{30_Bujanda_2017} Bujanda B, L{\'o}pez JL and Pagola PJ (2017) \textit{Convergent expansions of the incomplete gamma functions in terms of elementary functions}. Analysis and Applications \textbf{16}(3): 435--448. doi: \url{https://doi.org/10.1142/S0219530517500099}

\bibitem{31_Ansari_2019} Ansari IA (2019) \textit{The Analytical Solution of Incomplete Gamma Function to Determine the Electrical Resistivity at Normal State for $MgB_{2}$ Superconductor}. J. Phys. Conference Series \textbf{1172}: 012028. doi: \url{https://doi.org/10.1088/1742-6596/1172/1/012028}

\bibitem{32_Reynolds_2021} Reynolds R, Stauffer A (2021) \textit{A note on the summation of the incomplete gamma function}. Symmetry \textbf{13}(12): 2369. doi: \url{https://doi.org/10.3390/sym13122369}

\bibitem{33_Fejzullahu_2021} Fejzullahu BXh \textit{On the maximum value of a confluent hypergeometric function}. Comptes Rendus. Math{\'e}matique \textbf{359}(10): 1217--1224. doi: \url{https://doi.org/10.5802/crmath.256}

\bibitem{34_Pearson_2017} Pearson JW, Olver S and Porter MA (2017) \textit{Numerical methods for the computation of the confluent and Gauss hypergeometric functions}. Numerical Algorithm \textbf{74}(3): pages 821-–866. doi: \url{https://doi.org/10.1007/s11075-016-0173-0}

\bibitem{35_Bagci_2020} Ba{\u g}c{\i} A (2020) \textit{Advantages of Slater$-$type spinor orbitals in the Dirac$–$Hartree$–$Fock method. Results for hydrogen$-$like atoms with super$-$critical nuclear charge}. Rend. Fis. Acc. Lincei \textbf{31}(2): 369--385. doi: \url{https://doi.org/10.1007/s12210-020-00899-6}

\bibitem{36_Bagci_2022} Ba{\u g}c{\i} A (2022) \textit{JRAF: A Julia package for computation of relativistic molecular auxiliary functions}. Comput. Phys. Commun. \textbf{273}: 108276. doi: \url{https://doi.org/10.1016/j.cpc.2021.108276}

\bibitem{37_Klahn_1977} Klahn B, Bingel WA (1977) \textit{The convergence of the Rayleigh$-$Ritz Method in quantum chemistry}. Theoretica chimica acta, \textbf{44}(1): 27--43. doi: \url{https://doi.org/10.1007/BF00548027}

\bibitem{38_Magnus_1966} Magnus W, Oberhettinger F and Soni, RP (1966) \textit{Formulas and Theorems for the Special Functions of Mathematical Physics.} Springer-Verlag, New York

\bibitem{39_Lebedev_1965} Lebedev NN (1965) \textit{Special Functions and Their Applications.} Prentice-Hall Inc., London

\bibitem{40_Butera_2014} Butera S, Di Paola M (2014) \textit{A physically based connection between fractional calculus and fractal geometry}. Annals of Physics \textbf{350}: 146--158. doi: \url{https://doi.org/10.1016/j.aop.2014.07.008}

\bibitem{41_Liang_2019} Liang Y, Chen W and Cai W (2019) \textit{Hausdorff Calculus. Applications to Fractal Systems}. De Gruyter, Berlin.

\bibitem{42_Chen_2010} Chen Y, Yan Y and Zhang K (2010) \textit{On the local fractional derivative} J. Math. Anal App. \textbf{362}(1): 17--33. doi: \url{https://doi.org/10.1016/j.jmaa.2009.08.014}

\bibitem{43_Kolwankar_1997} Kolwankar KM, Gangal AD (1997) \textit{H{\"o}lder exponents of irregular signals and local fractional derivatives}. Pramana \textbf{48}(1): 49--68. doi: \url{https://doi.org/10.1007/BF02845622}

\bibitem{44_Khalil_2020} Khalil R, Al Horani M, Yousef A and Sababheh M (2014) \textit{A new definition of fractional derivative}. J. Comput. Appl. Math. \textbf{264}: 65--70. doi: \url{https://doi.org/10.1016/j.cam.2014.01.002}

\bibitem{45_Ortigueira_2019} Ortigueira MD, Martynyuk V, Fedula M and Machado JT (2019) \textit{The failure of certain fractional calculus operators in two physical models}. Fractional Calculus and Applied Analysis \textbf{22}(2): 255--270. doi: \url{https://doi.org/10.1515/fca-2019-0017}

\bibitem{46_Chung_2020} Chung WS, Zare S, Hassanabadi H and Maghsoodi E (2020) \textit{The effect of fractional calculus on the formation of quantum-mechanical operators}. Math. Method Appl. Sci. \textbf{43}(11): 6950--6967. doi: \url{https://doi.org/10.1002/mma.6445}

\bibitem{47_Ahmed_1999} El Sayed AMA (1999) \textit{On the generalized Laguerre polynomials of arbitrary (fractional) orders and quantum mechanics}. J. Phys. A Math. Gen. \textbf{32}(49): 8647--8654. doi: \url{https://doi.org/10.1088/0305-4470/32/49/305}

\bibitem{48_Ahmed_2000} El Sayed AMA (2000) \textit{Laguerre polynomials of arbitrary (fractional) orders}. App. Math. Comput. \textbf{109}(1): 1--9. doi: \url{https://doi.org/10.1016/S0096-3003(98)10112-1}

\bibitem{49_Ahmed_2004} Rida SZ, El Sayed AMA (2004) \textit{Fractional calculus and generalized Rodrigues formula}. App. Math. Comput. \textbf{147}(1): 29--43. doi: \url{https://doi.org/10.1016/S0096-3003(02)00648-3}

\bibitem{50_Mirevski_2010} Mirevski SP, Boyadjiev L (2010) \textit{On some fractional generalizations of the Laguerre polynomials and the Kummer function}. Comput. Math. with Appl. \textbf{59}(3): 1271--1277. doi: \url{https://doi.org/10.1016/j.camwa.2009.06.037}

\bibitem{51_Miana_2021} Miana PJ, Romero N (2021) \textit{Fractional Generalizations of Rodrigues-Type Formulas for Laguerre Functions in Function Spaces}. Mathematics \textbf{9}(9): 2227--7390. doi: \url{https://www.mdpi.com/2227-7390/9/9/984}

\bibitem{52_Bildstein_2018} Bildstein S (2018) \textit{Half theory fractional angular momentum and the application of fractional derivatives to quantum mechanics}. J. Math. Phys. \textbf{59}(2): 022110. doi: \url{https://doi.org/10.1063/1.4990102}

\bibitem{53_Bildstein_2018} Bildstein S (2018) \textit{Half theory. II. The application of fractional spherical harmonics to chemical bonding}. J. Math. Phys. \textbf{59}(8): 082101. doi: \url{https://doi.org/10.1063/1.5017744}

\bibitem{54_Jeng_2010} Jeng M, Xu SLY, Hawkins E and Schwarz JM (2010) \textit{On the nonlocality of the fractional Schrödinger equation} J. Math. Phys. \textbf{51}(6): 062102. doi: \url{https://doi.org/10.1063/1.3430552}

\bibitem{55_Dixit_2020} Dixit A, Ujlayan A (2020) \textit{Conformable Fractional Laguerre and Chebyshev Differential Equations with Corresponding Fractional Polynomials} in \textit{Computational Science and Its Applications}. Ed. by Siddiqi AH et.al. Chapman and Hall/CRC. doi: \url{https://doi.org/10.1201/9780429288739}

\bibitem{56_Atangana_2015} Atangana A, Baleanu D and Alsaedi A (2015) \textit{New properties of conformable derivative}. Open Mathematics \textbf{13}(1): 000010151520150081. doi: \url{https://doi.org/10.1515/math-2015-0081}

\bibitem{57_Shat_2019} Shat R, Alrefai S, Alhamayda I, Sarhan A and Al-Refai M (2019) \textit{The Fractional Laguerre Equation: Series Solutions and Fractional Laguerre Functions}. Front. Appl. Math. Stat. \textbf{5}: 2297--4687. doi: \url{https://doi.org/10.3389/fams.2019.00011}

\bibitem{58_Rasala_1981} Rasala R (1981) \textit{The Rodrigues formula and polynomial differential operators}. J. Math. Anal. Appl. \textbf{84}(2): 443--482. doi: \url{https://doi.org/10.1016/0022-247X(81)90180-3}

\bibitem{59_Weber_2007} Weber HJ (2007) \textit{Connections between real polynomial solutions of hypergeometric-type differential equations with Rodrigues formula}. Centr. Eur. J. Math. \textbf{5}(2): 415--427. doi: \url{https://doi.org/10.2478/s11533-007-0004-6}

\bibitem{60_Osler_1971} Osler TJ (1971) \textit{Fractional derivatives and Leibniz rule}. Amer. Math. Monthly \textbf{78}(6): 645--649. doi: \url{https://doi.org/10.2307/2316573}

\bibitem{61_Bhrawy_2016} Bhrawy AH, Taha TM, Abdelkawy MA, Hafez RM (2016) \textit{On Numerical Methods for Fractional Differential Equation on a Semi-infinite Interval} in \textit{Fractional Dynamics}. ED. by Cattani C et. al. De Gruyter Open, Warsaw, Poland. doi: \url{doi:10.1515/9783110472097-012}

\bibitem{62_Wang_2021} Wang WC (2021) \textit{Some notes on conformable fractional Sturm–Liouville problems}. Bound. Value Probl. \textbf{2021}(1): 103. doi: \url{https://doi.org/10.1186/s13661-021-01581-y} 

\end{thebibliography}
\end{document}